\documentclass[11pt,letterpaper]{article}
\usepackage{amsmath, amsthm, amssymb, wasysym, verbatim, bbm, bm, color, graphics, cancel, kotex}
\usepackage{geometry}
\usepackage{setspace}
\usepackage{lipsum}
\usepackage{ mathrsfs }
\usepackage[ruled, vlined]{algorithm2e}
\usepackage{lastpage}
\usepackage{enumerate}
\usepackage{fancyhdr}
\usepackage{natbib}
\usepackage{subcaption}
\usepackage[table,xcdraw]{xcolor}
\usepackage{graphicx, multirow}
\usepackage{listings}
\usepackage{ dsfont }
\usepackage{authblk}
\usepackage{tikz}
\usepackage[shortlabels]{enumitem}
\usepackage[T1]{fontenc}
\usepackage{tikz, pgfplots}
\usetikzlibrary{decorations,decorations.markings,decorations.text}
\usetikzlibrary{math}
\usepackage{tikz-dimline}
\usetikzlibrary{mindmap}
\usetikzlibrary{positioning}
\usetikzlibrary{arrows}
\usetikzlibrary{arrows.meta, positioning, shadows}
\tikzstyle{int}=[draw, fill=blue!20, minimum size=2em]
\tikzstyle{init} = [pin edge={to-,thin,black}]

\usepackage{hyperref}    %NOTE: needs to be defined as last package

\DeclareMathOperator*{\iid}{\overset{iid}{\sim}}

\newcommand{\R}{\mathbb{R}}

\def\bs{\boldsymbol}
\def\Ex{{\rm I\!E}}
\def\Pr{{\rm I\!P}}
\def\be{\begin{equation}}
	\def\ee{\end{equation}}
\def\bea{\begin{eqnarray*}}
	\def\eea{\end{eqnarray*}}
\def\bean{\begin{eqnarray}}
	\def\eean{\end{eqnarray}}
\def\nn{\nonumber}

\def\ra{\rightarrow}

\def\Bl{\Bigl}
\def\Br{\Bigr}

\def\alp{\alpha}

\def\del{\delta}
\def\eps{\epsilon}

\def\KK{{\mathcal K}}

\def\Yn{\mathbf{Y}\!_n}
\def\Zn{\mathbf{Z}_n}

\def\t{\mathbf{t}}
\def\Tt{T_{\t}(\Yn)}

\def\mm{\mathfrak{m}}

\newtheorem{thm}{Theorem}[section]

\newtheorem{prop}[thm]{Proposition}
\newtheorem{lem}[thm]{Lemma}

\theoremstyle{remark}

\SetKwInput{KwInput}{Input}                % Set the Input
\SetKwInput{KwOutput}{Output}              % set the Output
\title{Fast and Optimal Changepoint Detection and Localization using Bonferroni Triplets}
\author[1]{Jayoon Jang\thanks{Research supported by NSF grant DMS-1916074}} 
\author[1]{Guenther Walther\thanks{Research supported by NSF grants DMS-1916074 and DMS-2413885}}
\affil[1]{Stanford University}
\date{\today}

\begin{document}
	
\maketitle
	
\begin{abstract}
The paper considers the problem of detecting and localizing changepoints in a sequence of independent observations.
We propose to evaluate a local test statistic on a triplet of time points, for each such triplet in a particular collection.
This collection is sparse enough so that the results of the local tests can simply be combined with a weighted
Bonferroni correction. This results in a simple and fast method, {\sl Lean Bonferroni Changepoint detection} (LBD),
that provides finite sample guarantees for the existance of changepoints as well as simultaneous
confidence intervals for their locations. LBD is free of tuning parameters, and we
show that LBD allows optimal inference for the detection of changepoints. To this
end, we provide a lower bound for the critical constant that measures the difficulty of the changepoint detection problem,
and we show that LBD attains this critical constant. We illustrate LBD for a number of distributional settings, namely
when the observations are homoscedastic normal with known or unknown variance, for observations from a natural
exponential family, and in a nonparametric setting where we assume only exchangeability for segments without a changepoint.

\end{abstract}
	
\section{Introduction}

\subsection{Problem statement and motivation}

The detection and localization of changepoints in a sequence of observations is an important problem in statistics due
to its wide-ranging applications in areas such as finance, environmental monitoring, 
genomics, and signal processing; see e.g. the recent reviews
for those areas given in \cite{reeves2007review}, \cite{rybach2009audio},
\cite{zhang2010detecting}, and \cite{bardwell2019most}. In these applications, the motivation for accurately 
detecting changepoints 
is the desire to identify significant changes in the behavior of a system, with the ultimate goal
of  understanding the underlying causes driving those changes.

We investigate the case where one observes $n$ univariate independent observations $\Yn:=(Y_1, \ldots, Y_n)$. 
The integer
$\tau$ is called a changepoint if $Y_{\tau} \stackrel{d}{\neq} Y_{\tau +1}$.
We denote the set of changepoints as $\mathcal{T} = \{\tau_1, \ldots, \tau_K \}$, where
$\tau_0 := 0 < \tau_1 < \tau_2 < \cdots < \tau_K < \tau_{K+1} =: n$. 

Given the observations $Y_1,\ldots,Y_n$, the task is to detect and localize the changepoints while controlling 
the probability of a false discovery of a changepoint.
Our goal is to develop methodology that satisfies the following four criteria:

First, we want to give a finite sample guarantee for the false detection
probability that holds no matter what the true number of changepoints may be.
This is in contrast to most existing results in the literature where the guarantee is usually asymptotic 
and holds only in the case of no changepoint. 
Second, we want 
to give finite sample guarantees for the locations of the changepoints, i.e. we want to provide confidence intervals for these 
locations that
have a guaranteed finite sample coverage. 
Third, we show that the proposed methodology for the detection of changepoints is as powerful as possible in the sense
that it can detect all changepoints that are detectable. 
Fourth, we show that our methodology is computationally efficient with a complexity
of $O(n \log n)$. This is important in many applications, see e.g. the detection of DNA copy number variation discussed
in \cite{olshen2004circular, zhang2007modified}.

\subsection{Summary of key results} \label{summary}
	
We introduce \textit{Lean Bonferroni Changepoint Detection (LBD)} for detecting and localizing changepoints with 
finite sample guarantees. LBD produces a collection of intervals such that we can claim with guaranteed
finite sample confidence level $1-\alpha$ that every interval in the collection contains a changepoint.
 The key idea is to locally test for a changepoint at time $m$ using the observations on an interval $(s,e]\ni m$,
where the collection of triplets $(s,m,e)$ to be examined is 
sparse (`lean') enough so that the correspoding test statistics can simply be combined
with a weighted Bonferroni adjustment. An important point of this approach is the fact that it is possible
to construct such a collection of triplets that is sparse enough so that a simple Bonferroni
adjustment is not overly conservative, while at the same time the construction is sufficiently flexible so that
every triplet of changepoints $(\tau_{k-1},\tau_k,\tau_{k+1})$ can be approximated well enough in order to guarantee optimal changepoint detection.
An attendant benefit of the sparse construction is that it allows fast computation of LBD and no tuning
parameter is required.

We use local test statistics based on likelihood ratios and we specify these for a range of distributional settings, 
namely 
for the Gaussian case with known or unknown variance, for the case of a natural expontial family,
and for a nonparametric setting where one only assumes exchangeability if there are no changepoints.

`Optimal changepoint detection' means that LBD is able to detect every changepoint that is detectable
at all by any method. We make this precise in Section~\ref{power} for the prototypical Gaussian case
$$
                Y_i \ =\ \mu_i +\sigma Z_i, \quad i = 1, \ldots, n,
$$
where the mean vector $\bm{\mu} = (\mu_1, \ldots, \mu_n) \in \R^n$ is unknown, $\sigma >0$ is a known scale factor
and the $Z_i$ are i.i.d. N(0,1). 
Theorem~\ref{thm:detectgeneral} shows that 
LBD will detect with asymptotic power 1 a changepoint $\tau_k$ whose jump $|\mu_{\tau_k +1} -\mu_{\tau_k}|$ satisfies
\be \label{introbound}
|\mu_{\tau_k +1} -\mu_{\tau_k}| \, \sqrt{\frac{(\tau_k-\tau_{k-1})(\tau_{k+1}-\tau_k)}{\tau_{k+1}-\tau_{k-1}}}
\ \geq \ \left(\sqrt{2}+\eps_n(\tau_k)\right)
\sqrt{\log \frac{n}{(\tau_k-\tau_{k-1}) \wedge (\tau_{k+1}-\tau_k)}}
\ee
where $\eps_n(\tau_k)$ may go to zero at a rate of almost $\left(\log \frac{n}{(\tau_k-\tau_{k-1}) \wedge
(\tau_{k+1}-\tau_k)}\right)^{-\frac{1}{2}}$.
Theorem~\ref{thm:lb-single} presents a matching lower bound: It shows that if $\sqrt{2}+\eps_n(\tau)$ is replaced 
by $\sqrt{2}-\eps_n(\tau)$, then detection is impossible. 

We also give corresponding results for the simultaneous detection of a (potentially unbounded) number
of changepoints. In the special case where the distances between the changepoints are roughly equal, this 
results in the constant $\sqrt{2}$ being replaced by $2\sqrt{2}$ in (\ref{introbound}) and in the 
lower bound.

We point out that the key term for measuring the difficulty of this detection problem is the constant
$\sqrt{2}$ (resp. $2\sqrt{2}$) rather than the term $\sqrt{\log n}$. 
The reason for this is that this problem is related to
 the well known fact that the maximum
of $n$ i.i.d. standard normals concentrates around $\sqrt{ 2 \log n}$. Using the wrong constant 3 (say) in place
of 2 gives $\sqrt{ 3 \log n} =\sqrt{ 2 \log n^{3/2}}$ and thus pertains to the setting where the sample size
is $n^{3/2}$ instead of $n$. This heuristic explains why in detection problems, unlike in many other areas
of statistical inference, establishing optimality 
concerns the constant multiplier rather than the rate, see e.g. \cite{AriasCastroDonoho,walther2021calibrating2}. Therefore
Section~\ref{power} focuses on analyzing a neighborhood around this constant and gives a rate at which
$\eps_n(\tau) \ra 0$ in $\sqrt{2}\pm \eps_n(\tau)$. In contrast, results in the literature for various other
changepoint detection methods typically give a threshold of the form $C \sqrt{\log n}$ with a value for $C$ that can
be quite large, and it is therefore not clear whether these methods allow optimal inference.

We also point out that if one writes the optimal threshold as $C \sqrt{\log n}$, then the critical
constant $C$ depends on the distance between $\tau_k$ and the nearest changepoint,
i.e. $(\tau_k-\tau_{k-1}) \wedge (\tau_{k+1}-\tau_k)$. 
This is seen by writing $(\tau_k-\tau_{k-1}) \wedge (\tau_{k+1}-\tau_k) =n^s$
for some $ s\in [0,1)$, which shows that the RHS of (\ref{introbound}) 
equals $(\sqrt{2} \sqrt{1-s}+\eps_n(\tau)) \sqrt{\log n}$. Thus Theorem~\ref{thm:detectgeneral} shows
that LBD automatically attains the relevant critical constant no matter what the distance to the nearest
changepoint is.

There are two key attributes of LBD that result in optimal detection. The first is the particular construction
of the sparse collection of triplets on which the data are analyzed. This construction is a natural extension
of the Bonferroni intervals (2-tuples) introduced in \cite{walther2010optimal,walther2021calibrating}.
There are many proposals in literature for constructing approximating sets for intervals (2-tuples of points) 
that are both sparse (i.e. have close to $n$ intervals) and at the same time allow to
approximate any interval well. See e.g. \cite{AriasCastroDonoho} for an alternative construction. In fact,
\cite{kovacs2023seeded} generalize the construction introduced in \cite{walther2010optimal,rivera2013optimal} and call them `seeded
intervals'. However, a key motivation for the particular construction of \cite{walther2021calibrating} is 
that this specific construction allows to combine the
results on the various intervals with a simple weighted Bonferroni correction that allows optimal inference. 
Here we show how this construction
can be carried over to triplets of points. For this reason we will call these
triplets {\sl Bonferroni triplets}. The second key attribute of LBD is the weighting
employed for the Bonferroni correction. This weighting scheme allows to achieve optimal detection,
i.e. it achieves the critical constant irrespective of the distance to the nearest changepoint,
as explained above.

 The output of LBD is a collection $\mathcal{C}$ of confidence intervals that come with the finite sample
 guarantee that with simultaneous confidence $1-\alpha$ each interval $I \in \mathcal{C}$ contains a changepoint. 
This result is given in Theorem~\ref{thm:varKnownFinSample}. We note that localizing changepoints with
confidence intervals rather than with point estimates yields a more informative result due to the finite sample
guarantees on the location. Moreover, confidence intervals readily allow inference about other quantities that
may be of interest, and that inference will then inherit the finite sample guarantees.
For example, LBD immediately produces a lower confidence bound for the number of changepoints: 
If we can find $k$ disjoint confidence intervals for changepoints with simultaneous coverage
$1-\alpha$, then it follows
with confidence $1-\alpha$ that the number of changepoints must be at least $k$, hence $k$ is a 
$1-\alpha$ lower confidence bound for the number of changepoints, see Theorem~\ref{thm:numcpt}.

The collection $\mathcal{C}$ is fast to compute as the
 computational complexity of the method is $O(n)$ up to log terms. The nearly linear runtime is crucial for the analysis
of large data and is in contrast to many existing approaches.

\subsection{Literature survey and connections to existing work} \label{sec:past works}
	
There is a large literature on changepoint detection, see for example the overviews given in
\cite{niu2016multiple, truong2020selective,verzelen2023optimal}.
One common approach to changepoint detection employs a cost function 
\begin{align*}
		C(\mathcal{T}; Y_{1:n}) = \sum_{k=0}^K c(Y_{t_k+1}, \ldots, Y_{t_{k+1}}; \theta_k)
\end{align*}
where $c(\cdot)$ is a goodness-of-fit measure such as the negative log-likelihood, the $t_i$ are candidates
for changepoints, and the $\theta_k$ are relevant
parameters. A \textit{global optimization} approach minimizes the cost function with a 
penalty term that is usually proportional to the number of changepoints. Since the number of all possible segmentations of 
$\{1,\ldots,n\}$ is large, the global optimization problem can be computationally intensive. Pruned Exact Linear Time 
(PELT, \cite{killick2012optimal}) finds the exact solution when the penalty is linear in the number of changepoints. 
The complexity is $O(n)$ if the number of changepoints increases linearly with $n$ and the worst case complexity is $O(n^2)$. 
Simultaneous MUltiscale Changepoint Estimator (SMUCE \cite{frick2014multiscale}) minimizes the number of changepoints 
subject to  a multiscale goodness-of-fit constraint.  SMUCE also uses dynamic programming and has time complexity of 
$O(n^2)$ in general. 
	
A different line of research aims to obtain an \textit{approximate solution} to the above optimization problem. 
These methods sequentially find the best 
changepoint by local segmentation and stop according to some criterion. Binary Segmentation (BS, \cite{scott1974cluster}) 
uses a CUSUM statistic, which in that case
is a two-sample $z$-statistic for testing for a difference in the means of the samples $Y_{(t_1, t_2]}$ and 
$Y_{(t_2, t_3]}$, in order to greedily search for a changepoint. Starting with the full segment $(0, n]$, a maximizer w.r.t. $t_2$ of the 
CUSUM statistic is designated as a changepoint and the segment is split into two parts. The same routine is applied recursively
until a stopping criterion is satisfied. Since the recursive partitioning may result in some signals cancelling each
other in the test statistic, BS may miss changepoints and is thus not consistent for estimating changepoints.  
Wild Binary Segmentation(WBS, \cite{fryzlewicz2014wild}) recursively applies the CUSUM statistic on randomly chosen 
intervals. Seeded Binary Segmentation(SBS, \cite{kovacs2023seeded}) proposes to use a fixed set of seeded intervals rather than 
randomly sampled intervals. 

These works establish results on the detection of changepoints using assumptions on the \textit{minimum} of the jump sizes 
$\kappa_i = |\mu_{\tau_{i}+1} - \mu_{\tau_i}|$ and the \textit{minimum} of the distances to the nearest changepoint 
$\delta_i = \min(\tau_i - \tau_{i-1}, \tau_{i+1}- \tau_i)$. 
For instance, in WBS it is assumed for some sufficiently large $C > 0$ and any $\xi > 0$ 
\begin{align} \label{rate}
\left(\min_{i=1, \ldots, K} \kappa_i \right) \left(\sqrt{\min_{i=1, \ldots, K} \delta_i} \right) \ge C\log^{1/2+ \xi}(n)
\end{align} 
in order to prove the consistency of the algorithm, while Lemma~1 in \cite{Wang2020} shows that no 
consistent estimator exists if the LHS in (\ref{rate}) is smaller than $C\log^{1/2}(n)$.

A shortcoming of these results is that no guarantees are available if this condition is not met because
of the presence of changepoints that do not satify (\ref{rate}) , even if some other changepoints easily
satisfy this condition. 
%%In contrast, our condition (\ref{introbound}) guarantees detection irrespective of the existence of other changepoints that
%%may be too small to detect. 
Moreover, the condition (\ref{rate}) concerns the rate $\log^{1/2+ \xi}(n)$ and is thus
a rather lenient condition since the difficulty of the detection problem is characterized by the multiplier $C$ in $ C\log^{1/2}(n)$,
as was explained in \ref{summary}. 
%%In contrast, our condition (\ref{introbound}) and the corresponding lower bound address this
%%critical constant and thus establish optimality of our proposed method.

Finally, we point the reader to \cite{verzelen2023optimal}, who arguably present the most sophisticated theoretical analysis
for changepoint estimators to date. They analyze two new estimators: a least-squares estimator with a multiscale
penalty and a two-step post-processing procedure. In particular, \cite{verzelen2023optimal} establish a lower bound for
the detection of a fixed number of changepoints and they show that the above two estimators attain this bound apart
from a multiplicative constant, irrespective of the presence of other changepoints that may be too small to detect.
Since those results are up to a multiplicative constant, it is still an open problem whether these two procedures attain the optimal
critical constant. More importantly, the penalty term that is being used for these two methods is tailored to Gaussian
errors and optimality will be lost for non-Gaussian distributions, see the discussion in \cite{walther2021calibrating2}.
In contrast, the method introduced here adjusts for multiple testing by calibrating
significance levels rather than critical values. Therefore LBD readily allows for different error distributions
and is even applicable in a nonparametric setting, as will be demonstrated below. We show that LBD does in fact attain
the optimal critical constant in the setting with a fixed number of changepoints that is investigated in \cite{verzelen2023optimal}.
We also analyze a more general set-up, where the number of changepoints
is allowed to grow with the sample size. We believe that this is a more realistic setting for analyzing the detection of changepoints.
We establish a new lower bound for this setting which turns out to be different: The critical constant becomes twice as large
if one wishes to detect an increasing number of changepoints that are roughly equally spaced. We show that LBD attains the optimal
critical constant also in this more general setting. As in \cite{verzelen2023optimal}, the detection results for LBD also hold in the presence of
other changepoints that may be too small to detect. Finally, we point out that LBD does not involve any tuning parameter, whereas
most procedures in the literature require the user to set tuning parameters.

\section{Lean Bonferroni Changepoint Detection (LBD)}

In this section we introduce  our method  for detecting and localizing changepoints with finite sample
guarantees. This method also yields a lower confidence bound on the number of changepoints.
Section \ref{subsec:bonfInt}  describes Bonferroni triplets, which are a key building block for our method.
Section \ref{subsec:varKnown} explains how these Bonferroni triplets can be used for the simultaneous detection and 
localization of changepoints and for constructing a lower confidence bound on the number of changepoints. 
Section \ref{subsec:varUnknown} shows how the methodology can be applied in various distributional settings.

\subsection{Bonferroni triplets} \label{subsec:bonfInt}
	
Bonferroni triplets are a natural extension of {\sl Bonferroni intervals} introduced by 
\cite{walther2010optimal,walther2021calibrating}. Bonferroni intervals are
 a collection of $O(n \log n)$ intervals that provide a good approximation to the
$\sim n^2$ intervals $(j,k]$, $0 \leq j < k \leq n$. There are many ways to construct  a collection
of intervals that is both computationally efficient (i.e. has cardinality close to $n$) and which approximates
every interval $(j,k] \subset (0,n]$ well. 
For example, an alternative construction is given in \cite{AriasCastroDonoho}.
\cite{kovacs2023seeded} generalize the construction introduced in \cite{walther2010optimal,rivera2013optimal} and call them `seeded
intervals'. However, an important reason  for using the particular construction given in 
\cite{walther2021calibrating}
is that this specific construction allows to combine local tests
on the various intervals with a simple weighted Bonferroni correction such that the resulting inference is optimal.
This was shown by \cite{walther2021calibrating} for the problem of detecting a single interval with an elevated mean.
Here we will show  that this methodology can be extended from intervals (2-tuples of points) to pairs of abutting
intervals (triplets of points) in order to provide optimal inference about changepoints. This  demonstrates
that this methodology is useful in a more general context.

 In more detail, for a given integer $\ell \in \{0,\ldots,\ell_{max}:= \lfloor \log_2 (n/4) \rfloor -1\}$ we approximate intervals 
with lengths in $[2^{\ell}, 2^{\ell+1})$ by the collection 
$$
	\mathcal{J}_{\ell} := \Bigl\{ (j,k]: j,k \in \{id_{\ell}, i = 0, \ldots \}, 2^{\ell} \le k-j < 2^{\ell+1}  \Bigr\}
$$
The grid spacing $d_{\ell}$ is defined as $d_{\ell} = \lceil 2^{\ell}/\sqrt{2\log\frac{en}{2^{\ell}}} \rceil$. 
This spacing causes the 
relative approximation error to decrease with the interval length at a certain rate, see Lemma~\ref{approx}, 
which is required for optimal inference. The intervals in the collection $\bigcup_{\ell} \mathcal{J}_{\ell}$ are
called {\sl Bonferroni intervals}, see \cite{walther2021calibrating}. 
        
Now we extend these intervals by adjoining to the left or to the right
another interval whose length equals that of a Bonferroni interval that is at least as large:

We write $\mathcal{L}_n := \{\mbox{ lengths of Bonferroni intervals }\}$. Then the collection of 
{\sl Bonferroni triplets} is $\bigcup_{\ell=0}^{\ell_{max}} \mathcal{K}_{\ell}$, where
\begin{align*}
\mathcal{K}_{\ell} :=\ \Bigl\{ (t_1,t_2,t_3) :\ &(t_1,t_2] \in \mathcal{J}_{\ell} \mbox{ and }
t_3-t_2 \in \mathcal{L}_n,\   t_3-t_2 \geq t_2 -t_1,\ t_3 \leq n\\
\mbox{or }\ & (t_2,t_3] \in \mathcal{J}_{\ell} \mbox{ and }
t_2-t_1 \in \mathcal{L}_n ,\ t_2-t_1 > t_3 -t_2,\ t_1 \geq 0 \Bigr\}
\end{align*}
Note that the index $\ell$ refers to the size of the Bonferroni interval, whereas its extension may be much larger.
The reason for this indexing is that it turns out that in order to obtain optimal inference, the weights of the Bonferroni adjustment 
need to depend on the size index $\ell$ of
the Bonferroni interval but not on the size of its larger extension.
When working with the weighted Bonferroni adjustment given in Section~\ref{subsec:varKnown},
it is convenient to group certain triplets together. We select triplets whose Bonferroni intervals have sizes up to about $\log n$
and we call this group of triplets the first block, and we write 2nd, 3rd, $\ldots$ block for the subsequent $\mathcal{K}_{\ell}$:
\begin{align*}
	B \text{th block} := \begin{cases}
		\bigcup_{\ell=1}^{s_n-1} \mathcal{K}_{\ell} & B = 1 \\
		\mathcal{K}_{B-2+s_n} & B = 2, \ldots, B_{\max}
	\end{cases}
\end{align*}
where $s_n := \lceil \log_2 \log n \rceil$ and $\ B_{\max} := \lfloor \log_2 (n/4) \rfloor - s_n + 1$.

There are $O(n\log^{\frac{5}{2}} n)$ Bonferroni triplets in the collection $\bigcup_{\ell} \mathcal{K}_{\ell}$, which results
in a computational complexity of $O(n\log^{\frac{5}{2}} n)$ for the LBD methodology described below,
see Proposition~\ref{complexity}.

\subsection{Testing for a changepoint with LBD} \label{subsec:varKnown}

The definition of a changepoint makes it convenient to work with half-open intervals $(s,e]$.
For each Bonferroni triplet $\t=(s,m,e) $ we will evaluate a  local test statistic
$\Tt$ on $(Y_i, i \in (s,e])$ in order to test whether there is a changepoint at $m$.
Our methodology rests on two key points:
First, it is typically possible to give good tail bounds for the null distribution of $\Tt$. That is, if
there is no changepoint in  $(s,e)$, then we can give a good bound $c_{\t,n}(\cdot)$ for the
quantile function such that 
the following holds for $\alpha \in (0,1)$:
\be  \label{nulldist}
\mbox{If there is no changepoint in $(s,e)$, then for all $\t=(s,m,e)$: } 
\ \Pr \Bl(\Tt > c_{\t,n}(\alpha)\Br) \ \leq \ \alpha
\ee
In particular, the tail bound (\ref{nulldist}) holds whether there are changepoints outside $(s,e)$ or not.

In the second step, we  perform
simultaneous inference with the weighted Bonferroni adjustment that was introduced in \cite{walther2021calibrating}
for Bonferroni intervals: 
Each Bonferroni triplet $\t$ in the $B$th block is assigned the same
significance level $\alpha_{\t}$, and this $\alpha_{\t}$ is assigned weight $\frac{1}{B}$. Thus
the weight depends on the size of the Bonferroni interval in the triplet, but not on the size of its extension.
Hence for a given
simultaneous confidence level $1-\alpha$ we set
$$
\alpha_{\t}\ :=\ \dfrac{\alpha}{B \,(\sum_{b=1}^{B_{max}} b^{-1})\ \# (B \text{th block})} \quad \quad
\mbox{ where $B$ is the block index of $\t$.}  
$$

The reason for using this particular weighted Bonferroni correction is that it results in optimal inference irrespective
of the distance between changepoints,
when used in conjuction with appropriate local test statistics $\Tt$.
This  will be shown in Section~\ref{power}.

{\sl Lean Bonferroni Changepoint Detection (LBD)} declares  with finite sample confidence $1-\alpha$ that there 
is a changepoint in every interval $(s,e)$ for which a Bonferroni triplet $\t=(s,m,e)$ results
in $\Tt > c_{\t,n}(\alpha_{\t})$. We
write $\mathcal{C}(\alp)$ for the collection of the resulting intervals:
$$
\mathcal{C}(\alp):= \Bl\{J:\  J=[s+1,e-1] \mbox{ and } \Tt >
c_{\t,n}(\alp_{\t}) \mbox{ for some Bonferroni triplet } \t=(s,m,e)\Br\}
$$

It follows from Bonferroni's inequality and (\ref{nulldist}) that
$\Pr\bigl( LBD \mbox{ erroneously declares a changepoint} \bigr)  \leq \alpha$, 
regardless of how many detectable or non-detectable changepoints there are in $(Y_1,\ldots,Y_n)$,
see Theorem~\ref{thm:varKnownFinSample}.
Therefore the intervals $J \in \mathcal{C}(\alp)$ are confidence intervals for the locations of changepoints with 
simultaneous finite sample confidence level at least $1-\alp$. 

It is typically sufficient to work with the intervals $J \in \mathcal{C}(\alp)$ that are minimal with respect
to inclusion, i.e. for which no proper subset exists in $\mathcal{C}(\alp)$. This is because if a changepoint
results in a significant interval $(s,e)$, then this changepoint may also cause a number of supersets
of $(s,e)$ to be significant.
But these supersets are typically not of interest since they result in less precision for localizing the changepoint.
It is straightforward to compute the minimal
intervals in $\mathcal{C}(\alp)$, see Algorithm~\ref{alg:maxDisjoint} in Appendix~\ref{a:maxDisjoint}.

The collection $\mathcal{C}(\alp)$ can be used for a number
of inferential tasks.
Of particular interest is 
$$
N(\alp)\ :=\ \mbox{largest number of disjoint intervals in $\mathcal{C}(\alp)$}
$$
If each interval $J \in \mathcal{C}(\alp)$ contains a changepoint, then the existence
of $N(\alp)$ disjoint intervals $J \in \mathcal{C}(\alp)$ implies that $N(\alp)$ cannot be
larger than the number of changepoints in $(Y_1,\ldots,Y_n)$. Since the former condition holds with
confidence $1-\alp$,  
it follows that $N(\alp)$ is a $(1-\alp)$ lower confidence bound for the number
of changepoints in $(Y_1,\ldots,Y_n)$. No nontrivial upper confidence bound can exist, see \cite{donoho1988one}.
We summarize these results in the following theorem:

\begin{thm} \label{thm:varKnownFinSample}
For every $n \geq 2$:
$$
\Pr \Bigl(\mbox{ each interval $J \in \mathcal{C}(\alp)$ contains a changepoint} \Br)\ \geq \ 1-\alp
$$
Let $K$ denote the number of changepoints in $(Y_1,\ldots,Y_n)$. Then
$$
\Pr \Bigl( N(\alp) \leq K \Bigr) \ \geq \ 1-\alp
$$
\end{thm}

\subsection{Applying LBD in various distributional settings} \label{subsec:varUnknown}

In order to apply LBD it remains to specify the test statistic $\Tt$. We will show that in parametric settings
the generalized likelihood ratio statistic results in optimal inference.
We will also present a nonparametric $\Tt$
that is applicable when no parametric specification for the the distribution of the $Y_i$ is appropriate.

The two requirements for the test statistic are that it be as powerful as possible and that it satisfies
the pivotal tail bound (\ref{nulldist}). To this end, we will employ the square-root of twice the log generalized
likelihood ratio statistic. 
This statistic has also been used in other detection problems, see e.g.
\cite{rivera2013optimal,frick2014multiscale,Konig,walther2021calibrating}.
In more detail,  let $\t=(s,m,e)$ be a Bonferroni triplet. Then we can define 
the test statistic 
\be  \label{GLR}
\Tt:=\sqrt{2 \log \text{LR}_{\t}(\Yn)}
\ee
where $\text{LR}_{\t}(\Yn)$ denotes the generalized likelihood ratio statistic for testing no changepoint in $(s,e)$ 
vs. a changepoint at $m$, based on $(Y_i, i \in (s,e])$.  Hence if $Y_i \sim f_{\theta_i}$ for some parametric family 
$f_{\theta}$, then
\be  \label{LR}
 \text{LR}_{\t}(\Yn) \ =\  \dfrac{\Bl(\sup_{\theta \in \Theta} \prod_{i= s+1}^m f_{\theta}(Y_i)\Br) \ 
\Bl(\sup_{\theta \in \Theta} \prod_{i= m+1}^{e} f_{\theta}(Y_i)\Br)}{\sup_{\theta \in 
  \Theta}\prod_{i=s+1}^e f_{\theta}(Y_i)}
\ee
The reason for using the transformation $\sqrt{2 \log \text{LR}_{\t}(\Yn)}$
is that it produces a statistic whose null distribution is close to the absolute value of a standard normal.
More importantly, this transformation admits good finite sample tail bounds, see \cite{walther2022tail}.
Hence this statistic is nearly pivotal and satisfies property (\ref{nulldist}) with simple tail bounds
that do not even depend on $\t$, as will be seen below.

We now illustrate LBD for various distributional settings. We start with the simple Gaussian model
\be \label{model}
                Y_i \ =\ \mu_i +\sigma Z_i, \quad i = 1, \ldots, n,
\ee
where the scale factor $\sigma >0$ is known and the $Z_i$ are i.i.d. N(0,1). The interest in
this model arises from the fact that optimality results derived for the Gaussian model can be expected to carry over
to other distributional settings, see \cite{GramaNussbaum,BrownLow}. 
For this reason we will focus on this model when we establish
optimality results for LBD in Section~\ref{power}. A standard calculation shows that (\ref{GLR}) results in
\be  \label{localtest}
\Tt\ =\ \frac{\big| \overline{Y}_{(s,m]} -\overline{Y}_{(m,e]}\big|}{\sigma }
   \sqrt{\frac{(m-s)(e-m)}{e-s}}
\ee
where $\overline{Y}_{(s,m]}:=\frac{1}{m-s}\sum_{i=s+1}^e Y_i$.
Thus $\Tt$ equals a local two-sample z-statistic.
Now (\ref{model}) immediately yields:
$$
\mbox{If $\bm{\mu}$ is constant on $(s,e]$, then } \Tt =T_{\t}(\Zn) \sim  \big|\mbox{N(0,1)} \big|
$$

Thus the tail bound (\ref{nulldist}) holds with
\be  \label{eq:critVal}
c_{\t,n}(\alpha) := z\left(\dfrac{\alpha}{2} \right)
\ee
where $z(\alpha)$ denotes the $(1-\alpha)$-quantile of $N(0,1)$.

If $\sigma$ is unknown in the Gaussian model (\ref{model}), then a straightforward calculation shows that
the statistic $\sqrt{2 \log \text{LR}_{\t}}$ is equivalent to the two-sample $t$-statistic
\be  \label{two-t}
\Tt\ :=\ \frac{\big|\overline{Y}_{(s,m]} -\overline{Y}_{(m,e]}|}{s_p }
   \sqrt{\frac{(m-s)(e-m)}{e-s}}
\ee
where $s_p^2=\frac{1}{e-s-2} \left(\sum_{i=s+1 }^m(Y_i-\overline{Y}_{(s,m]})^2 + \sum_{i=m+1}^e (Y_i -
\overline{Y}_{(m,e]})^2 \right)$ is the pooled sample variance\footnote{Since we need at least four observations 
for estimating the
pooled sample variance $s_p^2$, we will in this case employ only Bonferroni triplets whose Bonferroni intervals have length at least 2.}.
Note that if $\bm{\mu}$ is constant on $(s,e]$, then  $\Tt \sim t_{e-s-2}$. Hence the tail bound (\ref{nulldist})
holds with the $t_{e-s-2}$-quantile in place of the standard normal quantile $z(\cdot)$ in (\ref{eq:critVal}).

As another example, let $Y_i \sim f_{\theta_i}$, $i = 1, \ldots, n$ be independent observations from a regular  
one-dimensional natural exponential family $\{f_{\theta}, \theta \in \Theta\}$ with density 
$f_{\theta}(x)= \exp(\theta x - A(\theta)) h(x)$ with respect to some $\sigma$-finite measure.
Then (\ref{LR}) gives
\be  \label{logLR}
\log \text{LR}_{\t}(\Yn)\ =\ \sup_{\theta \in \Theta} \Bl( \theta \sum_{i= s+1}^m Y_i -(m-s) A(\theta)\Br) +
\sup_{\theta \in \Theta} \Bl( \theta \sum_{i= m+1}^e Y_i -(e-m) A(\theta)\Br) -
\sup_{\theta \in \Theta} \Bl( \theta \sum_{i= s+1}^e Y_i -(e-s) A(\theta)\Br) 
\ee

For example, if the $Y_i$ are Poisson with natural parameter $\theta_i= \log \Ex Y_i$, then one computes
$$
\sqrt{2 \log \text{LR}_{\t}(\Yn)}\ =\ \sqrt{2  (m-s) \overline{Y}_{(s,m]} \log 
\frac{\overline{Y}_{(s,m]}}{\overline{Y}_{(s,e]}} +
2(e-m) \overline{Y}_{(m,e]} \log \frac{\overline{Y}_{(m,e]}}{\overline{Y}_{(s,e]}} }
$$
while letting the $Y_i$ have the exponential distribution with natural parameter $\theta_i= - (\Ex Y_i)^{-1}$ gives
\be \label{expdist}
\sqrt{2 \log \text{LR}_{\t}(\Yn)}\ =\ \sqrt{2 (m-s) \log \frac{ \overline{Y}_{(s,e]}}{\overline{Y}_{(s,m]}} +
2(e-m) \log \frac{ \overline{Y}_{(s,e]}}{\overline{Y}_{(m,e]}}}
\ee

Again it turns out that this transformation admits good finite sample tail bounds, see \cite{walther2022tail}:

If $\theta_i$ is constant on $(s,e]$, then for all $x>0$
$$
\Pr \left(  \sqrt{2\log \text{LR}_{\t}(\Yn)} > x \right)  \ \le \ (4+2x^2) \exp \left(-\frac{x^2}{2}\right)
$$
and in fact the better bound  $(4+2e)\exp(-x^2/2)$ holds if $x$ is not very large.
Using  the latter bound one sees that (\ref{nulldist}) holds for natural exponential families with 
$$
c_{\t,n}(\alpha)\ :=\ \sqrt{2 \log \frac{4+2e}{\alpha}}
$$

LBD can also be applied in a nonparametric setting.
If one does not want to make any assumption on the distribution of the $Y_i$, then an appropriate set-up would be
to require only that the $(Y_i,\, i \in (s,e])$ are exchangeable if there is no changepoint in $(s,e]$. This setting can
be analyzed with local rank statistics: Let $(R_i^{(s,e]}, \, i \in (s,e])$ be ranks of the $(Y_i,\, i \in (s,e])$, so
$(R_i^{(s,e]}, \, i \in (s,e])$ is permutation of $(1,2,\ldots, e-s)$. If one evaluates the statistic $T_{\t}$ in (\ref{localtest})
with the $R_i^{(s,e]}$ in place of the $Y_i$, then one obtains the Wilcoxon rank-sum statistic
$\frac{1}{m-s} \sum_{i=s+1}^m R_i^{(s,e]} -\frac{e-s+1}{2}$, apart from a scaling factor.
This statistic is sensitive to a changepoint at $m$ that manifests as a change in location.

If there is no changepoint in $(s,e]$, then $(R_i^{(s,e]}, \, i \in (s,e])$ is a random permutation of $(1,2,\ldots, e-s)$
and exact p-values can be computed if $e-s$ is not large. For large $e-s$ the normal approximation can be used, or if one
wishes to work with guranteed but somewhat conservative p-values, then
Corollary~6.6 of \cite{Duembgen2002} gives the finite sample tail bound
$$
\Pr \left( \Tt:=\sqrt{\frac{12 (m-s)}{(e-s+1)^2}} \ \left| \frac{1}{m-s} \sum_{i=s+1}^m R_i^{(s,e]} 
-\frac{e-s+1}{2} \right| >x \right) \ \leq \ 2 \exp\left(-\frac{x^2}{2} \right)
$$
for all $x>0$. Thus (\ref{nulldist}) holds again with
$$      
c_{\t,n}(\alpha)\ :=\ \sqrt{2 \log \frac{2}{\alpha}}
$$

Finally, we point out that in this case the $(1-\alpha_{\t})$-quantile of $\Tt$ can easily be simulated.
This is also true for some, but not for all, exponential families. For example, if the $Y_i \sim \text{Exp}(\theta)$,
then one readily checks that $\sqrt{2 \log \text{LR}_{\t}(\Yn)}$ in (\ref{expdist}) is a function of a $F_{2(m-s),2(e-m)}$
random variable. One potential disadvantage of simulation is that a large number of simulation runs may be required in order
to control the multiple simulation errors of the $c_{\t,n}(\alpha_{\t})$. While the $c_{\t,n}(\cdot)$ are identical for
many $\t$ because of invariance, there may be $\sim (\log n)^3$ different quantile functions $c_{\t,n}(\cdot)$
as there are $\sim (\log n)^{\frac{3}{2}}$ different lengths of Bonferroni intervals. 
This shows the advantage of having tight tail bounds, which are made possible by the $\sqrt{2 \log \text{LR}_{\t}}$ transformation.

\subsection{LBD allows fast computation}

In the parametric cases considered above it is possible to compute $\Tt$ for a given Bonferroni triplet $\t$ 
in $O(1)$ steps by pre-computing the cumulative sums of the $Y_i$ and the $Y_i^2$.
Hence the sparsity of the collection of Bonferroni triplets is the key for a fast computation of LBD.

\begin{prop} \label{complexity}
There are $O(n \log^{\frac{5}{2}} n)$ Bonferroni triplets in $\bigcup_{\ell} \mathcal{K}_{\ell}$, and
LBD can be computed in $O(n \log^{\frac{5}{2}} n)$ steps when the data are from an exponential family.
\end{prop}

\section{LBD achieves optimal detection}  \label{power}

In this section we investigate the power properties of LBD. We focus on the Gaussian model (\ref{model})
with known standard deviation $\sigma$, which without loss of generality we may assume to be 1:
$$
Y_i=\mu_i +Z_i, \qquad i=1,\ldots,n
$$
where the $Z_i$ are i.i.d. standard normal. Focusing on this case is motivated by the fact that
optimality results derived for this Gaussian model can be expected to carry over
to other distributional settings, see \cite{GramaNussbaum,BrownLow}. 

$K_n$ denotes the number of changepoints of $\bs{\mu}_n$.
Our goal is the simultaneous detection of $m_n$ of these $K_n$ changepoints, where $1 \leq m_n \leq K_n$.
So $m_n=1$ concerns the detection of one particular changepoint, whereas $m_n=K_n$ deals with the detection
of all changepoints. 
We restrict ourselves to the practically most relevant situation where the distances to the next
changepoints are small relative to $n$: All results
are derived under the assumption that each of these $m_n$ changepoints $\tau_k$ satisfies
$(\tau_k -\tau_{k-1}) \vee (\tau_{k+1}-\tau_k) \leq n^p$
for some $p \in (0,1)$. As explained following Theorem ~\ref{thm:lb-single}, the results can be extended
to larger interchangepoint distances with some small modifications.

It turns out that the key condition for the simultaneous detection of $m_n $ changepoints is that each 
of these $m_n$ changepoints satisfies
\be  \label{eq:st}
|\mu_{\tau_k+1} -\mu_{\tau_k}| \, \sqrt{\frac{(\tau_k-\tau_{k-1})(\tau_{k+1}-\tau_k)}{\tau_{k+1}-\tau_{k-1}}}
\ \geq \  \sqrt{2\log \frac{n}{(\tau_k-\tau_{k-1}) \wedge (\tau_{k+1}-\tau_k)}} + \sqrt{2\log m_n} +b_n
\ee
where $b_n \uparrow \infty$ is any sequence (common to the $m_n$ changepoints) that may diverge arbitrarily slowly,
see Theorem~\ref{thm:detectgeneral} below.
We point out that this result allows for the simultaneous detection of changepoints with unevenly spaced jumps,
 where some of these changepoints
satisfy (\ref{eq:st}) by having a large jump and a small distance to the next changepoint, while other changepoints
satisfy (\ref{eq:st}) by having a small jump and a large distance. Importantly, this result also guarantees
detection in the presence of other changepoints that do not satisfy this condition and which
therefore may not be detected. 

Condition (\ref{eq:st})  has the following interpretation: The term 
$\sqrt{2\log \frac{n}{(\tau_k-\tau_{k-1}) \wedge (\tau_{k+1}-\tau_k)}}$ accounts for the multiple
testing for changepoints in $(0,n]$, so this term is a consequence of controlling the type I error. The term
$\sqrt{2\log m_n}$ is necessary in order to ensure that all changepoints are actually detected, so this term
is a consequence of simultaneously controlling the type II error. The term $b_n$, 
which can be negligible compared to the other two terms,
is required in order to guarantee asymptotic power 1 in the aforementioned detection.
	
It is informative to interpret this condition for some special cases:
	
In the case where one is concerned about one particular changepoint or a bounded number $m_n$ of changepoints,
then condition (\ref{eq:st}) becomes
\be  \label{low1}
|\mu_{\tau_k +1} -\mu_{\tau_k}| \, \sqrt{\frac{(\tau_k-\tau_{k-1})(\tau_{k+1}-\tau_k)}{\tau_{k+1}-\tau_{k-1}}}
\ \geq \ \left(\sqrt{2}+\eps_n(\tau_k)\right) 
\sqrt{\log \frac{n}{(\tau_k-\tau_{k-1}) \wedge (\tau_{k+1}-\tau_k)}}
\ee
where $\eps_n(\tau_k)$ may go to zero at a rate of almost $\left(\log \frac{n}{(\tau_k-\tau_{k-1}) \wedge 
(\tau_{k+1}-\tau_k)}\right)^{-\frac{1}{2}}$.
	
On the other hand, if one is concerned about the detection of all changepoints and the $\tau_{k+1}-\tau_k$ are of
similar size, then $m_n \sim \frac{n}{\tau_{k+1}-\tau_k}$ and condition (\ref{eq:st}) becomes
\be  \label{low2}
|\mu_{\tau_k +1} -\mu_{\tau_k}| \, \sqrt{\frac{(\tau_k-\tau_{k-1})(\tau_{k+1}-\tau_k)}{\tau_{k+1}-\tau_{k-1}}}
\ \geq \ \left(2\sqrt{2}+\eps_n(\tau_k)\right)
\sqrt{\log \frac{n}{(\tau_k-\tau_{k-1}) \wedge (\tau_{k+1}-\tau_k)}}
\ee
where $\eps_n(\tau)$ may go to zero at a rate of almost $\left(\log \frac{n}{(\tau_k-\tau_{k-1}) \wedge 
(\tau_{k+1}-\tau_k)}\right)^{-\frac{1}{2}}$.
	
If a changepoint is detectable, then we can ask about how precisely we can localize it. 
We say that a confidence interval $C(\alp)$ localizes a changepoint $\tau$ with precision $\delta$
if  $\tau \in C(\alp) \subset [\tau -\del, \tau +\del]$. 
Equivalently, the precision is $\sup_{x\in C(\alp)} |x-\tau|$,
provided $\tau \in C(\alp)$.
In particular, precision $\delta<1$ implies
that that the confidence interval contains the single integer $\tau$ and thus localizes $\tau$ exactly.
We  point out that the literature typically addresses the localization of changepoints with point estimates
rather than with confidence intervals, see e.g. \cite{verzelen2023optimal},
and the resulting precisions will not be the same. For example, if we take the midpoint
of $C(\alp)$ as an estimator $\hat{\tau}$ of a changepoint, then $\tau \in C(\alp) \subset [\tau -\del, \tau +\del]$
implies $|\hat{\tau}- \tau| \leq \del/2$. Thus the precision of the point estimate $\hat{\tau}$
is better by at least a factor of 2, but a point estimate does not come with a confidence interval for the
location.

%------------------------------Minor note: ----------------------------
%If $T_I$, $I=(s,e]$, is significant, then can claim that there is a changepoint in $[s+1,e-1]$, so precision is
%not worse than $e-s-1=|I|-1$
%(even: not worse than anything larger than $|I|-2$).
%If $\tau$ is a changepoint and $T_I$, $I=(\tau -\delta,\tau+\delta]$, is significant, then
%precision is $\delta=|I|/2$ (or anything larger than $\delta -1$.)
%---------------------------------------------------------------------

In order to heuristically explain our result about localizing $\tau_k$ given in Theorem~\ref{thm:detectgeneral} below, 
note that by (\ref{eq:st}) $\tau_k$ is detectable provided
that $|\mu_{\tau_k +1} -\mu_{\tau_k}| \sqrt{
\frac{(\tau_k-\tau_{k-1})(\tau_{k+1}-\tau_k)}{\tau_{k+1}-\tau_{k-1}}}$ is large enough; \cite{verzelen2023optimal} call
this product the `energy' of $\tau_k$. Now we observe that $\frac{(\tau_k-\tau_{k-1})(\tau_{k+1}-\tau_k)}{\tau_{k+1}-\tau_{k-1}}$ is roughly
proportional to the distance between $\tau_k$ and the nearest changepoint, i.e. roughly proportional to
the number of observations available to test whether $\tau_k$ is a changepoint. Hence if the jump size 
$|\mu_{\tau_k +1} -\mu_{\tau_k}| $ is so large that there is slack in the inequality (\ref{eq:st}), then the changepoint
would still satisfy (\ref{eq:st}) with a smaller value of $\frac{(\tau_k-\tau_{k-1})(\tau_{k+1}-\tau_k)}{\tau_{k+1}-\tau_{k-1}}$, 
i.e. with fewer observations between $\tau_{k-1}$ and $\tau_{k+1}$. 
Therefore we expect to be able to localize $\tau_k$ among these fewer observations around $\tau_k$  and obtain a corresponding precision.
Indeed, Theorem~\ref{thm:detectgeneral} shows that LBD detects each changepoint $\tau_k$ that satisfies (\ref{eq:st})
and moreover localizes it in the smallest interval on which the energy exceeds the threshold given in (\ref{eq:st}):

\begin{thm} \label{thm:detectgeneral}
Let $\bs{\mu}_n \in \R^n$ be arbitrary mean vector.
Let $\tau_{k_1},\ldots,\tau_{k_{m_n}}$ be those changepoints of $\bs{\mu}_n:=(\mu_1,\ldots,\mu_n)$ that satisfy
(\ref{eq:st}).
Then
$$
\Pr_{\bs{\mu}_n} \Bigl(\mbox{ LBD detects all $\tau_{k_1},\ldots,\tau_{k_{m_n}}$
		and localizes each $\tau_{k_i}$  with precision given in (\ref{precision}}\Bigr)
\ \ra \ 1
$$
uniformly in $\bs{\mu}_n$. 

Set $g(x):= \left( \frac{\sqrt{2\log \frac{n}{x}} + \sqrt{2\log m_n} +b_n}{
\mu_{\tau_{k_i}+1} -\mu_{\tau_{k_i}}}\right)^2$ and $\mm_i:=\min(\tau_{k_i}-\tau_{k_i-1}, \tau_{k_i+1}-\tau_{k_i})$.
Then the precision is not worse than
\be  \label{precision}
\begin{cases}
2 g\left(\left( \mu_{\tau_{k_i}+1} -\mu_{\tau_{k_i}}\right)^{-2} \right) & \text{if (\ref{eq:st}) holds with $\mm_i$ in place
of both $\tau_{k_i}-\tau_{k_i-1}$ and $\tau_{k_i+1}-\tau_{k_i}$,}\\
& \text{ and with $2\mm_i$ in place of $\tau_{k_i+1}-\tau_{k_i-1}$}\\
\left( 1-\frac{g(\mm_i)}{\mm_i}\right)^{-1} g(\mm_i) & \text{otherwise.}
\end{cases}
\ee
\end{thm}

Theorem~\ref{thm:lb-single} below gives a lower bound for detecting a single changepoint and
this lower bound matches the constant $\sqrt{2}$ in (\ref{low1}), while
Theorem \ref{thm:lb-simul} gives a lower bound for simultaneously detecting  equally spaced changepoints and that lower bound
matches the constant $2\sqrt{2}$ in (\ref{low2}). 
Together with Theorem~\ref{thm:detectgeneral}, these results show
that the critical constant for detecting one (or a bounded number of) changepoints  is $\sqrt{2}$, the critical constant
for simultaneously detecting equally spaced changepoints is $2\sqrt{2}$, and that LBD attains
these critical constants and is therefore optimal for the detection of changepoints. Note that the detection condition
(\ref{eq:st}) for LBD covers the general case of unequally spaced changepoints; we use equally spaced changepoints for our lower bounds
in order to keep the exposition simple, but an inspection of the proofs shows that these can be readily modified
for unequally spaced changepoints.

\begin{thm} \label{thm:lb-single}
Let $\bs{\mu}_n \in \R^n$ be an arbitrary mean vector that may change with $n$. The only requirement on  
$\bs{\mu}_n$ is that the number of its changepoints $K_n$ is not larger than $\frac{n}{4}$.
		
Given such a vector $\bs{\mu}_n$ we can change $\bs{\mu}_n$ locally to obtain $v_n$ vectors 
$\bs{\mu}_n^{(i)} \in \R^n$, $i=1,\ldots,v_n$, where $v_n \in [K_n, \frac{n}{4}]$ is any divergent sequence, 
such that each $\bs{\mu}_n^{(i)} $ has all the changepoints that $\bs{\mu}_n$ has and at least one additional 
changepoint $\tau_{k_i}^{(i)}$ that satisfies
\be  \label{1a}
\left|\mu_{\tau_{k_i}^{(i)} +1}^{(i)} -\mu_{\tau_{k_i}^{(i)}}^{(i)}\right| \, 
\sqrt{\frac{\left(\tau_{k_i}^{(i)}-\tau_{k_i-1}^{(i)}\right)\left(\tau_{k_i+1}^{(i)}-
\tau_{k_i}^{(i)}\right)}{\tau_{k_i+1}^{(i)}-\tau_{k_i-1}^{(i)}}}\ \geq \ 
\left(\sqrt{2}-\epsilon_n\right) \sqrt{\log \frac{n}{\left(\tau_{k_i}^{(i)}-
\tau_{k_i-1}^{(i)}\right) \wedge \left(\tau_{k_i+1}^{(i)}-\tau_{k_i}^{(i)}\right)}}
\ee
where $\eps_n=\eps_n(\tau_{k_i}^{(i)}) \ra 0$ satisfies 
$\eps_n(\tau_{k_i}^{(i)})\sqrt{\log\frac{n}{\left(\tau_{k_i}^{(i)}-\tau_{k_i-1}^{(i)}\right) \wedge
\left(\tau_{k_i+1}^{(i)}-\tau_{k_i}^{(i)}\right)}} \to \infty$, and the following holds:
		
If $\psi_n (\bs{Y}_n)$ is any level $\alp$ test under $\bs{\mu}_n$ (e.g. a test of $\bs{\mu}_n$ against additional
changepoints) then
$$
\min_{i=1,\ldots,v_n} \Pr_{\bs{\mu}_n^{(i)}} \Bl( \psi_n (\bs{Y}_n) \mbox{ rejects }\Br) \ \leq \ \alp +o(1).
$$
\end{thm}
	
Note that the lower bound given in Theorem~\ref{thm:lb-single} is based on a `local analysis': It shows that
it is impossible to detect an additional changepoint for a {\sl given} vector $\bs{\mu}_n$. In contrast, the
impossibility result in \cite{verzelen2023optimal} considers the detection of a changepoint against a zero background.
Theorem~\ref{thm:lb-single} also provides a sharper bound of the form $\sqrt{2}-\epsilon_n$, while \cite{verzelen2023optimal}
establish impossibility for $\sqrt{2-\xi}$ for every $\xi \in (0,2)$.

As an aside we note that if one is interested in large $\del (t_i) \sim n$, then a contiguity argument as in 
Theorem~4.1.(c) in
\cite{dumbgen2008multiscale} shows that (\ref{eq:st}) is necessary for the detection of $\tau$ with asymptotic
power 1. Conversely, LBD will detect $\tau$ with asymptotic power 1 if
$b_n$ is replaced by $\log \log n$ in (\ref{eq:st}). 
	
The critical constant for detecting a single point clearly also holds for the simultaneous detection of a bounded 
number of changepoints. In the more relevant case where one is interested in a number $m_n$ of changepoints that 
grows with $n$, however, simultaneous detection requires in general that the critical constant is twice as large, 
as is shown by Theorems \ref{thm:detectgeneral} and \ref{thm:lb-simul}.

\begin{thm} \label{thm:lb-simul}
Let $m_n \le n/4$ be a divergent sequence of positive integers. Then there exists a mean vector $\bs{\mu}_n \in \R^n$  
whose changepoints contain $\tau_{k_i}, i = 1, \ldots, m_n$, each of which satisfies \be  \label{eq:lb-simulreq}
\left|\mu_{\tau_{k_i} +1} -\mu_{\tau_{k_i}}\right| \,
\sqrt{\frac{\left(\tau_{k_i}-\tau_{k_i-1}\right)\left(\tau_{k_i+1}-
\tau_{k_i}\right)}{\tau_{k_i+1}-\tau_{k_i-1}}}\ \geq \
\left(2\sqrt{2}-\epsilon_n\right) \sqrt{\log \frac{n}{\left(\tau_{k_i}-
\tau_{k_i-1}\right) \wedge \left(\tau_{k_i+1}-\tau_{k_i}\right)}}
\ee
where $\eps_n=\eps_n(\tau_{k_i}) \ra 0$ satisfies 
$\eps_n(\tau_{k_i})\sqrt{\log\frac{n}{\left(\tau_{k_i}-\tau_{k_i-1}\right) \wedge
\left(\tau_{k_i+1}-\tau_{k_i}\right)}} \to \infty$, and the following holds:
		
If $\psi_n (\bs{Y}_n)$ is any procedure for detecting changepoints with significance level $\alpha$, then
$$
\Pr_{\bs{\mu}_n} \Bl( \psi_n (\bs{Y}_n) \mbox{ detects all changepoints } \tau_{k_i}, i = 1, \ldots, m_n \Br) \ = \ \alp +o(1).
$$
\end{thm}
	
The precision (\ref{precision}) equals, up to a multiplicative constant, the localization error of the point estimates given
in \cite{verzelen2023optimal}, who also show that this localization error matches, up to a multiplicative constant, a lower bound.
As in the case of detection, we submit that the difficulty of the localization problem is reflected by the multiplicative constant
rather than by the rate, see Section~\ref{summary}. 
We leave it as an open problem to establish this critical constant for point estimates and for 
confidence intervals.

Finally, we point out that if $\tau_{k+1}-\tau_k =o(\tau_k -\tau_{k-1})$, then the detection condition (\ref{low1}) is equivalent
to the condition (2.2) given in \cite{chanwalther} for the detection of a segment $(\tau_k,\tau_{k+1}]$ of elevated means against a zero background.
This `segment detection problem' has been extensively studied in the literature as it is the prototypical model for the use
of the scan statistic.
The results in this section show that segment detection is not intrinsically easier than changepoint detection, thus answering
an open question posed in \cite{verzelen2}[Section 6.3].
	
\subsection{Inference about the number of changepoints}

It is known that the appropriate way to perform inference about the number of changepoints
of $(Y_1,\ldots,Y_n)$ is via a lower confidence bound. The reason why no nontrivial upper confidence
bound or no meaningful point estimate can exist 
is because there are $(\widetilde{Y}_1,\ldots,\widetilde{Y}_n)$ 
that are empirically indistinguishable from $(Y_1,\ldots,Y_n)$ but with the distributions of the 
$\widetilde{Y}_i$ having
a changepoint at every $i$. Therefore the coverage probability of an upper confidence bound cannot be guaranteed
but will necessarily depend on the unknown jump sizes and distances between the changepoints.
See e.g. \citep{donoho1988one} for an exposition in a general context.

There is another relevant conceptual issue concerning the inference about the number $K=K_n$ of changepoints:
Even if a procedure is available that detects all $K_n$ changepoints in an optimal way, the resulting lower
confidence bound for $K_n$ may be far from $K_n$. This is because there is generally
some uncertainty in the localization of the detected changepoints, and this uncertainty may prevent
the conclusion that two detected changepoints are distinct. We illustrate this in the particular
case of LBD: Suppose that each of the $K_n$ changepoints $\tau_i$ satisfies the detection condition (\ref{eq:st})
without slack. Then LBD detects each $\tau_i$ and the corresponding minimal confidence interval is close
to $(\tau_{i-1},\tau_{i+1})$, as can be seen from the proof of Theorem~\ref{thm:detectgeneral}.
So we can claim with confidence $1-\alpha$ that there is a changepoint in each interval $(\tau_{i-1},\tau_{i+1})$, 
$i=1,\ldots,K_n$. However, these intervals overlap for consecutive indices $i$ and $i+1$,
and therefore we cannot conclude that the corresponding changepoints are distinct.
Counting disjoint significant intervals as in Section~\ref{subsec:varKnown}
will therefore result in a  confidence bound $N(\alp)$ for $K_n$ that is about
$K_n/2$ rather than a tight bound equal to $K_n$. 

The next theorem shows that increasing the critical constant in (\ref{eq:st}) by a factor of $\sqrt{2}$ will 
guarantee that LBD will give a tight lower confidence bound $N(\alp)$ that equals $K_n$, apart from 
the noncoverage probability at most $\alp$:

\begin{thm} \label{thm:numcpt}
Let $\bs{\mu}_n \in \R^n$ be an arbitrary mean vector. Let $\tau_1, \ldots, \tau_{K_n}$ be 
changepoints of $\bs{\mu}_n$, each of which satisfies
\be \label{detectdisjoint}
|\mu_{\tau_k+1} -\mu_{\tau_k}| \, \sqrt{\frac{(\tau_k-\tau_{k-1})(\tau_{k+1}-\tau_k)}{\tau_{k+1}-\tau_{k-1}}}
\ \geq \  \sqrt{4\log \frac{n}{(\tau_k-\tau_{k-1}) \wedge (\tau_{k+1}-\tau_k)}} + \sqrt{4\log m_n} +b_n
\ee     
where $b_n \uparrow \infty$ is any sequence that may diverge arbitrarily slowly.
Then 
$$
\Pr_{\bs{\mu}_n} \Bigl(\mbox{LBD will produce $K_n$ disjoint confidence intervals for changepoints
}\Bigr) \ \ra \ 1
$$
uniformly in $\bs{\mu}_n$. Therefore
$$
\liminf_{n \ra \infty} \Pr_{\bs{\mu}_n} \Bigl( N(\alp)=K_n \Bigr) \ \geq \ 1-\alp
$$
uniformly in $\bs{\mu}_n$.
\end{thm}

\section{Numerical experiments} \label{sec:num}

We examine the coverage of the collection $\mathcal{C}(\alpha)$
of simultaneous confidence intervals and of the lower bound $N(\alp)$ for
 the number of change points on simulated data. Code implementing LBD, reproducing
 the experiments and data analysis is available at \url{https://github.com/Jayoon/LBD}.

%\subsection{Changepoint detection on simulated data}
%\textcolor{blue}{This section shows that the method has the claimed guarantees by simulated data. We also compare the method with several methods such as WBS, SBS, SMUCE etc.}

\subsection{Setup} \label{subsec: setup} 
\begin{figure}[ht]
        \begin{subfigure}{.5\textwidth}
                \centering
                \includegraphics[width=.8\linewidth]{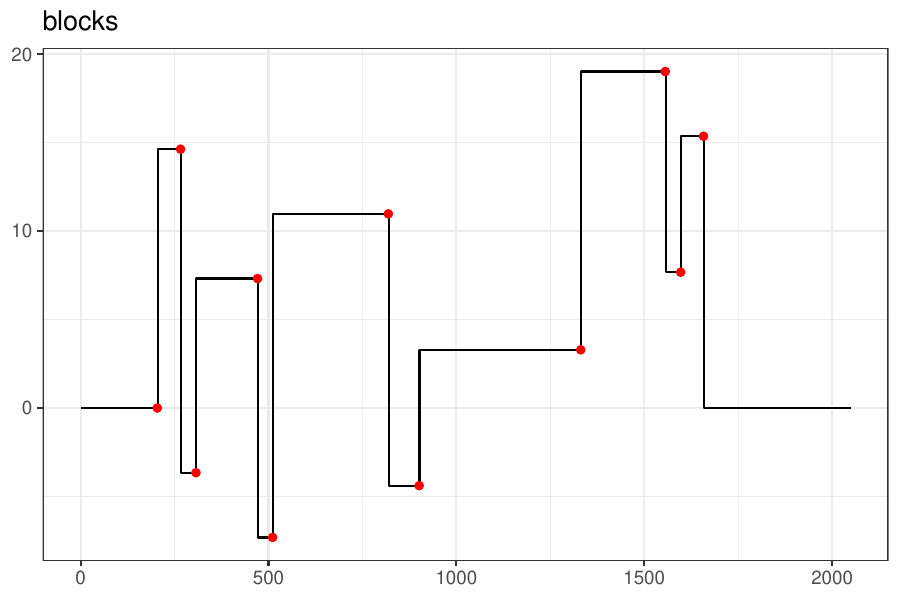}
        \end{subfigure}%
        \begin{subfigure}{.5\textwidth}
                \centering
                \includegraphics[width=.8\linewidth]{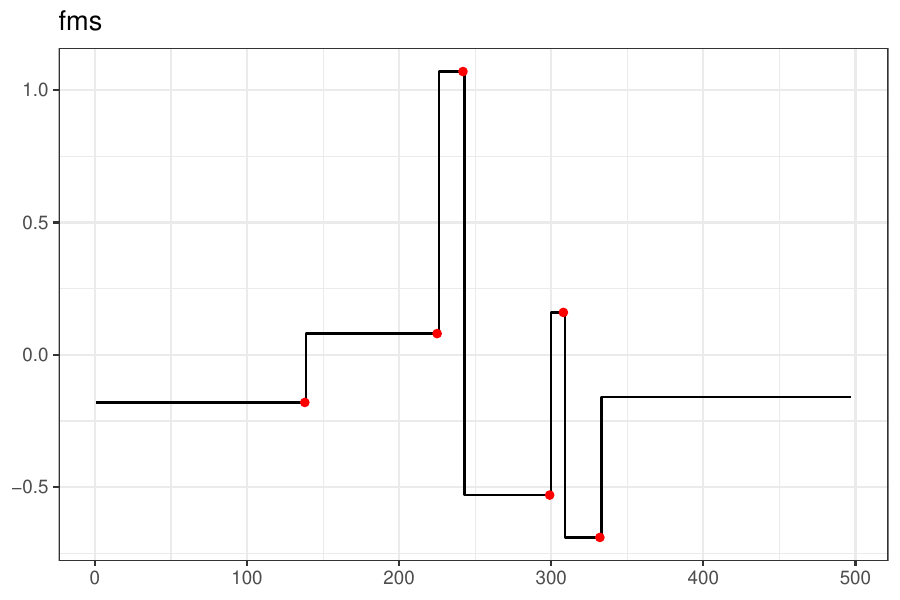}
                %\caption{1b}
                %\label{fig:sfig2}
        \end{subfigure}
        
        \begin{subfigure}{.5\textwidth}
                \centering
                \includegraphics[width=.8\linewidth]{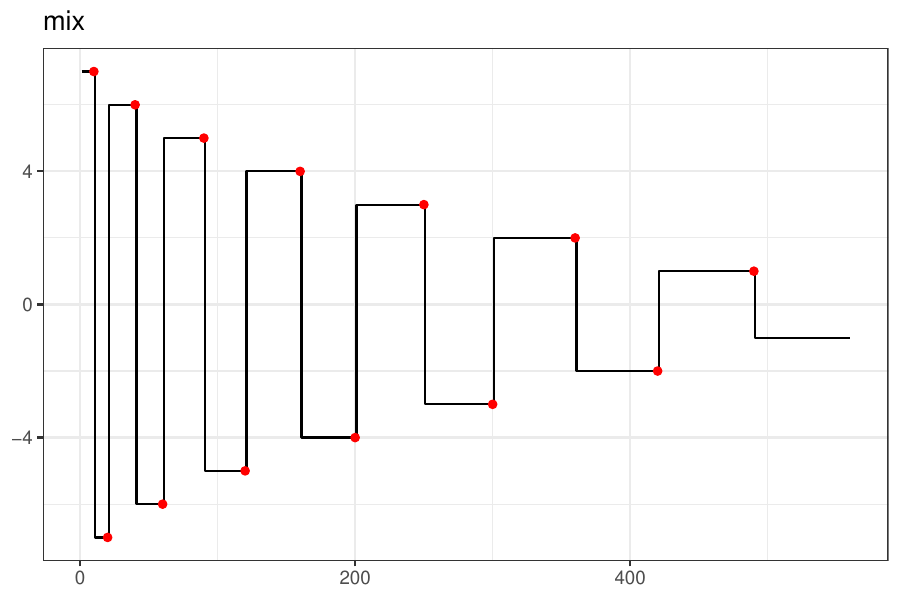}
                %\caption{1a}
                %\label{fig:sfig1}
        \end{subfigure}%
        \begin{subfigure}{.5\textwidth}
                \centering
                \includegraphics[width=.8\linewidth]{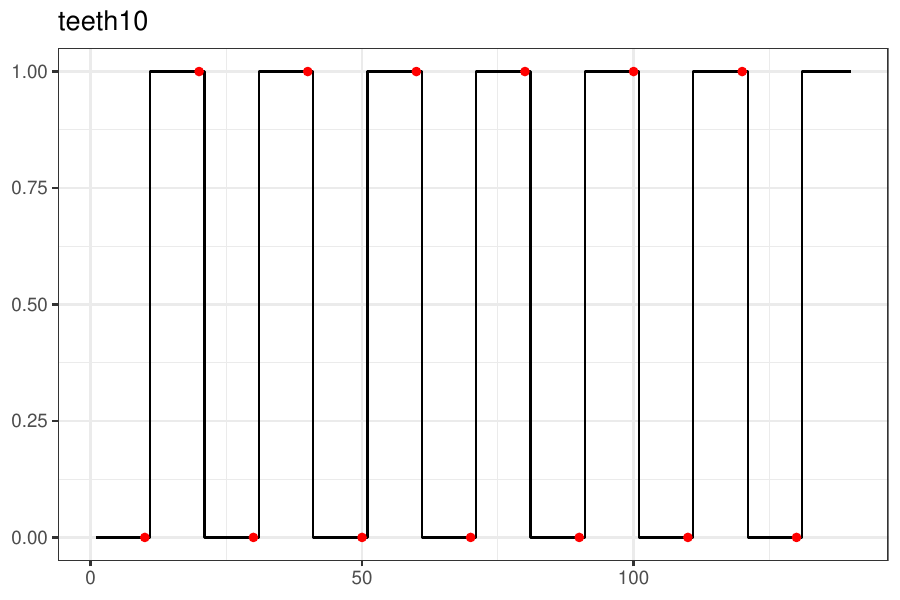}
                %\caption{1b}
                %\label{fig:sfig2}
        \end{subfigure}
        
        \begin{subfigure}{.5\textwidth}
                \centering
                \includegraphics[width=.8\linewidth]{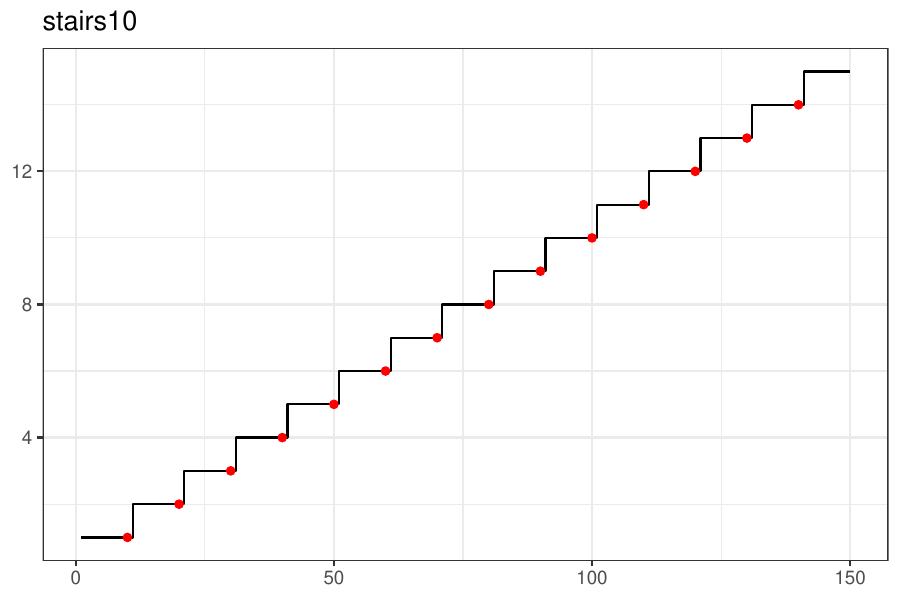}
                %\caption{1a}
                %\label{fig:sfig1}
        \end{subfigure}%
        \begin{subfigure}{.5\textwidth}
                \centering
                \includegraphics[width=.8\linewidth]{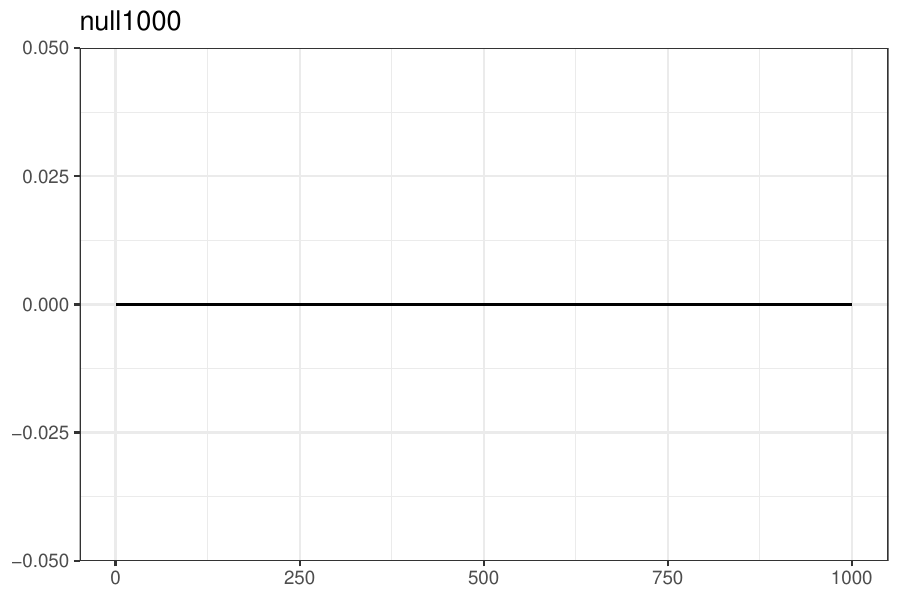}
                %\caption{1a}
                %\label{fig:sfig1}
        \end{subfigure}
        
        \caption{Signal functions $\bm{\mu}$ with changepoints marked as red dots.}
        \label{fig:signals}
\end{figure}

We generate $n$ i.i.d.~samples $Y_1, \ldots, Y_n$ with $Y_i = \mu_i + \sigma Z_i$ where the $Z_i$ are i.i.d. 
$\mathcal{N}(0,1)$. The signal functions $\bm{\mu}$ and the standard deviations $\sigma$ are selected from a set commonly used in the changepoint 
detection literature, see Figure~\ref{fig:signals} and Appendix~\ref{a.simul.fun}.

\subsection{Results}

We apply LBD with significance level $\alpha = 0.1$ and obtain the collection $\mathcal{C}(\alpha)$
 of simultaneous confidence sets.
Then we apply Algorithm~\ref{alg:maxDisjoint} to get the lower confidence bound $N(\alpha)$. 
For each setting, we generate a total of $n_{\text{sim}} = 10,000$ datasets and we compute the average of the
resulting $N(\alpha)$. We validate Theorem~\ref{thm:varKnownFinSample} by computing 
\begin{align*}
        \hat{p}_1= \dfrac{1}{n_{\text{sim}}} \sum_{i=1}^{n_{\text{sim}}} I\Bigl(\mbox{ each interval $J \in 
\mathcal{C}_i(\alp)$ contains a changepoint of $\bm{\mu}$} \Bigr)
\end{align*} 
and 
\begin{align*}
  \hat{p}_2 = \dfrac{1}{n_{\text{sim}}} \sum_{i=1}^{n_{\text{sim}}} I\Bigl(N_i(\alpha) \le 
\text{number of changepoints in $\bm{\mu}$}\Bigr)
\end{align*} 
where $\mathcal{C}_i(\alpha)$ and $N_i(\alpha)$ refer to the $i$th simulated dataset. 
Table \ref{tbl:results} shows the result for each signal function. 

\begin{table}[htbp]
\centering
\begin{tabular}{lr|lrrrr
>{\columncolor[HTML]{EFEFEF}}r rl|rrr}
signals   & \multicolumn{1}{l|}{$K$} & $\le -5$     & -4    & -3    & -2     & -1   & 0   & 1  & \multicolumn{1}{r|}{2}     & 
     \multicolumn{1}{l}{$\overline{N_i(\alpha)}$} & \multicolumn{1}{l}{$\hat{p}_1$} & \multicolumn{1}{l}{$\hat{p}_2$} \\ \hline
null 1000 & 0  &  & \multicolumn{1}{l}{} & \multicolumn{1}{l}{} & \multicolumn{1}{l}{} & \multicolumn{1}{l}{} & \textbf{0.987}  & 0.013 & 
     \multicolumn{1}{r|}{0.000} & 0.013   & 1.000     & {0.987}   \\
null 2000 & 0  &  & \multicolumn{1}{l}{} & \multicolumn{1}{l}{} & \multicolumn{1}{l}{} & \multicolumn{1}{l}{} & \textbf{0.990}   & 
     0.010  & \multicolumn{1}{r|}{0.000} & 0.011   & 1.000    & {0.990}   \\
null 3000 & 0  &  & \multicolumn{1}{l}{} & \multicolumn{1}{l}{} & \multicolumn{1}{l}{} & \multicolumn{1}{l}{} & \textbf{0.987}   & 
     0.013  & \multicolumn{1}{r|}{0.000} & 0.013  & 1.000  & 0.987    \\
blocks    & 11  & \multicolumn{1}{r}{0.009} & 0.109 & 0.376  & \textbf{0.387}  & 0.117  & 0.001  & \multicolumn{1}{l}{} &  & 
     8.499   & 0.993  & 1.000     \\
fms  & 6   &    & \multicolumn{1}{l}{} & 0.017  & 0.212  & \textbf{0.583}  & 0.187   & 0.001 &   & 4.943 & 0.992  & 0.999   \\
mix  & 13  & \multicolumn{1}{r}{0.008} & 0.078  & 0.376  & \textbf{0.455}  & 0.082   & 0.002 & \multicolumn{1}{l}{} &  & 10.529  & 
     0.995   & 1.000     \\
teeth 10  & 13 & \multicolumn{1}{r}{0.442} & \textbf{0.272}  & 0.184  & 0.080   & 0.019 & 0.003  & \multicolumn{1}{l}{} &   & 8.685  & 
     0.996  & 1.000   \\
stairs 10 & 14  &  & 0.001   & 0.009  & 0.100  & 0.401  & \textbf{0.488} & 0.001  &   & 13.371   & 0.996   & 0.999
\end{tabular}
\caption{Relative frequencies of the differences between the lower bound $N(\alpha)$ and $K$,  
average of the $N_i(\alpha)$, $\hat{p}_1$ and $\hat{p}_2$, for various signals. 
$K$ denotes the true number of changepoints in the signal. The most frequent entry for each signal is 
highlighted in bold.}
\label{tbl:results}
\end{table}

We can see that $\hat{p}_1$ and $\hat{p}_2$ are all greater than $1-\alpha = 0.9$, which is expected from the 
result of Theorem~\ref{thm:varKnownFinSample}. Furthermore, the observed coverage sometimes well exceeds $0.9$.

\section{Real Data Example}
A common application of changepoint detection algorithms is in the analysis of copy number variations (CNVs), 
where the objective is to identify changes in DNA copy numbers within a cell line, see
 \cite{zhang2010detecting, hyun2021post}. For this study, we utilized array comparative genomic hybridization 
(aCGH) data from \cite{snijders2001assembly}, which quantifies expression levels across the genome as the 
logarithm of the ratio in fluorescence intensity between test and reference samples. Additionally, external 
spectral karyotyping results are available for this dataset, providing a reference point for evaluating the output 
of our changepoint detection algorithm.

Since the normality assumption is questionable for these data,
we applied the LBD method with the Wilcoxon rank-sum statistic and exact p-values, setting the significance level $\alpha$ to 
0.05 and focusing specifically on the gm05296 cell line. In total, we analyzed 2116 measurements distributed across 
23 chromosomes. Since the measurements were normalized to be centered around zero, we opted to apply the method to 
the entire chromosome rather than conducting separate analyses for each chromosome. This also helped circumvent 
potential issues related to multiple testing.

\begin{figure}[h]
\centering
\includegraphics[width=\textwidth]{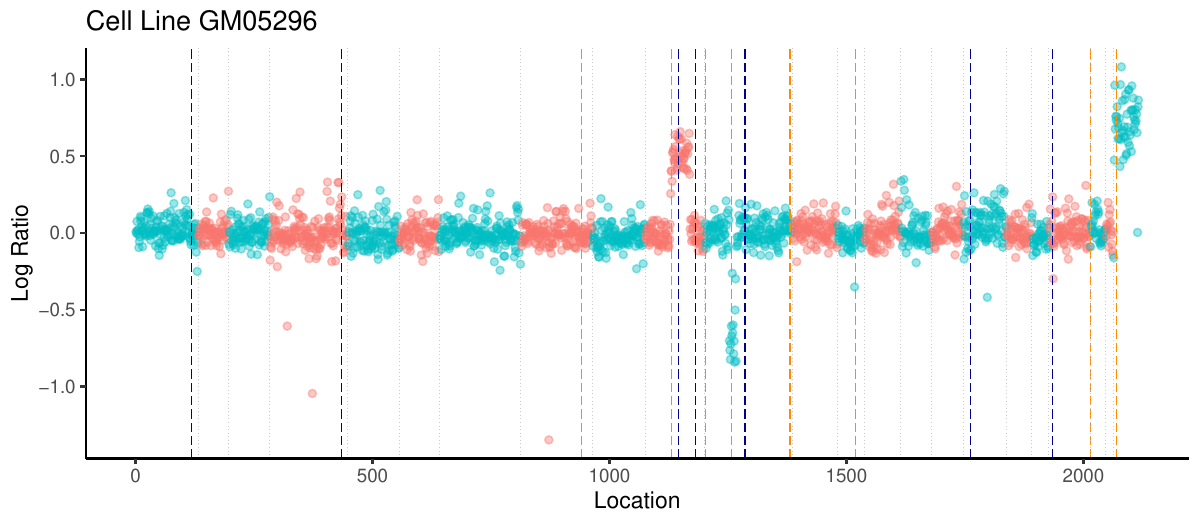}
\caption{Normalized log ratio plot of cell line gm05296 across 23 chromosomes from 
 \cite{snijders2001assembly}. Chromosome boundaries are demarcated by alternating green and red segments, 
 complemented by grey dotted lines for enhanced visibility. Additionally, eight disjoint confidence intervals are 
 highlighted with vertical dashed lines in alternating colors (blue and orange).}
\label{fig:real_data}
\end{figure}

Figure \ref{fig:real_data} visually represents the data and highlights the disjoint confidence intervals. 
We identified a total of 32 minimal intervals, including 8 disjoint intervals. The second and third intervals  
correspond to changepoints on chromosome 10, and the fourth and fifth intervals correspond to changepoints  chromosome 11. 
These changepoints align with external karyotyping results.
The detections corresponding to the other four intervals lack corresponding karyotyping data. These instances 
may represent false positives, or alternatively, genuine copy number variations that were not discernible 
through karyotyping. In fact, there is a notable and clearly visible jump on chromosome 23.

\bibliographystyle{apalike}
\bibliography{scan}

\appendix
\section{Computing the minimal intervals and the largest number of disjoint intervals in $\mathcal{C}(\alp)$} 
\label{a:maxDisjoint}

Given the collection $\mathcal{C}(\alp)$ of intervals, Algorithm~\ref{alg:maxDisjoint} computes the collection
$\mathcal{M}(\alp)$ of minimal (w.r.t. inclusion) intervals of $\mathcal{C}(\alp)$, i.e. $\mathcal{M}(\alp)$
contains those intervals $J \in \mathcal{C}(\alp)$ for which no proper subset exists in $\mathcal{C}(\alp)$. 
Algorithm~\ref{alg:maxDisjoint} also computes $N(\alp)$, the largest number of disjoint intervals in $\mathcal{C}(\alp)$,
as well as a collection $\mathcal{D}(\alp)$ that contains $N(\alp)$ disjoint intervals from $\mathcal{C}(\alp)$.
The intervals in $\mathcal{D}(\alp)$ are also minimal. The intervals in $\mathcal{C}(\alp)$ need not be distinct,
but those in $\mathcal{M}(\alp)$ and in $\mathcal{D}(\alp)$ will be.

\begin{algorithm} 
\KwInput{Collection of intervals $\mathcal{C}(\alp)$ }
\textbf{Process: }
\begin{enumerate}
\item $\mathcal{M}(\alp) = \emptyset, \ \mathcal{D}(\alp) = \emptyset, \ f = -\inf,\ g=-\inf,\ h=-\inf$
\item Sort the intervals in $\mathcal{C}(\alp)$ in increasing order w.r.t the right endpoints. For intervals
having the same right endpoint: sort in decreasing order w.r.t the left endpoint. This results in a list of intervals
$[a_1, b_1], \ldots, [a_m,b_m]$, with $m:=\# \mathcal{C}(\alp)$.
\item {\bf for} $i=1,\ldots,m$:
\begin{itemize}
\item {\bf if} $a_i > f$ {\bf then} \ $\mathcal{D}(\alp) \leftarrow \mathcal{D}(\alp) \cup \{[a_i,b_i]\}, \ f\leftarrow b_i$
\item {\bf if} $(a_i >g \mbox{ and } b_i>h)$ {\bf then} \ $\mathcal{M}(\alp) \leftarrow \mathcal{M}(\alp) \cup 
\{[a_i,b_i]\}, \ g\leftarrow a_i,  \ h\leftarrow b_i$
\end{itemize}
\end{enumerate}
\KwOutput{$\mathcal{M}(\alp), \mathcal{D}(\alp), N(\alp):=\# \mathcal{D}(\alp)$}
\caption{Computes $\mathcal{M}(\alp)$, $\mathcal{D}(\alp)$, $N(\alp)$}
\label{alg:maxDisjoint}
\end{algorithm}  
        
The time complexity of this algorithm is $O(m \log m)$ due to the sorting required in step 2. 

\paragraph{Proof of the correctness of Algorithm~\ref{alg:maxDisjoint}} 

First we show that if $[a_i,b_i]$ is minimal, then $[a_i,b_i] \in \mathcal{M}(\alp)$:

Since the $b_i$ are non-decreasing, we must have $b_i \geq h$ every time the if-condition is evaluated in step 3 of
the algorithm. If $b_i=h$, then $h>-\infty$, hence an interval $[a_k,b_k]=[g,h]$ was added to $\mathcal{M}(\alp)$
in an earlier iteration of  the for-loop, so $k<i$. But $b_k=b_i$ and $k<i$ imply $a_k \geq a_i$ due to the
sorting of the intervals. Therefore either $[a_k,b_k]=[a_i,b_i]$ in which case $[a_i,b_i] \in \mathcal{M}(\alp)$
as was to be shown, or $[a_k,b_k]\subsetneq [a_i,b_i]$. But that latter relation is a contradiction to
$[a_i,b_i]$ being minimal, so we need only consider the case $b_i >h$ from here on.
If $a_i \leq g$, then $g>-\infty$, so there is a previously added interval $[g,h] \subsetneq [a_i,b_i]$,
contradicting the minimality of $[a_i,b_i]$. Thus $a_i >g$ and $b_i >h$, which implies that 
$[a_i,b_i]$ will be added to $ \mathcal{M}(\alp)$ in step 3 of the algorithm.
\smallskip

Next we show that every $[a_i,b_i] \in \mathcal{M}(\alp)$ is minimal:

Let $[a_i,b_i] \in \mathcal{M}(\alp)$ and suppose that $[a_i,b_i]$ is not minimal. Then there exists
a minimal $[a_k,b_k] \subsetneq [a_i,b_i]$. This implies $k<i$ due to the sorting of the intervals, as well
as $[a_k,b_k] \in \mathcal{M}(\alp)$ by the first part of the proof.
When adding $[a_k,b_k]$ to $\mathcal{M}(\alp)$, step 3 of the algorithm sets $g=a_k$. Since $k<i$ and since
updates to $g$ can only increase $g$, the inequality $g\geq a_k \geq a_i$ holds when the algorithm evaluates
the if-condition for $[a_i,b_i]$. But this means that $[a_i,b_i]$ will
{\sl not} be added to $\mathcal{M}(\alp)$ in step 3 of the algorithm, which is a contradiction to our assumption.
Hence $[a_i,b_i]$ must be minimal.
\smallskip

Next we show that the collection $\mathcal{D}(\alp)$ contains $N(\alp)$ disjoint intervals:

Since the algorithm initializes $f=-\inf$, it follows immediately from step 3 that
\be  \label{algo3}
[a_1,b_1] \mbox{ is the first interval that is added to $\mathcal{D}(\alp)$}
\ee
Since the $b_i$ are non-decreasing, the condition $a_i>f$ requires that the left endpoint of $[a_i,b_i]$
is larger than the right endpoint of the interval that was previously added to $\mathcal{D}(\alp)$.
It follows that
\be \label{algo1}
\mbox{the intervals in $\mathcal{D}(\alp)$ are disjoint}
\ee
and therefore
\be  \label{algo2}
\# \mathcal{D}(\alp) \ \leq \ N(\alp)
\ee

If $N(\alp)=1$, then (\ref{algo3}) and (\ref{algo2}) imply $\# \mathcal{D}(\alp)=1$, which proves the claim.

If $N(\alp) \geq 2$, then by the definition of $N:=N(\alp)$ there exist intervals
$$
[a_{s(1)},b_{s(1)}],\, [a_{s(2)},b_{s(2)}], \ldots, [a_{s(N)},b_{s(N)}] \in  \mathcal{C}(\alp)
$$
such that
\be  \label{algo4}
b_{s(j)} < a_{s(j+1)},\quad j=1,\ldots,N-1
\ee
Denote the intervals that the algorithm successively adds to $\mathcal{D}(\alp)$ by
$$
[a_{t(1)},b_{t(1)}],\, [a_{t(2)},b_{t(2)}], \ldots
$$
We will show by induction for $j=1,\ldots,N$:
\be  \label{IS}
\mbox{The algorithm adds a $j$th interval $[a_{t(j)},b_{t(j)}]$ to $\mathcal{D}(\alp)$
and $b_{t(j)} \leq b_{s(j)}$}
\ee
(\ref{algo3}) shows that $[a_1,b_1]$ is added as the first interval, so $t(1)=1$. Hence (\ref{IS})
obtains for $j=1$ since $b_{t(1)} =b_1 \leq b_{s(1)}$ as $b_1$ is the smallest $b_i$.

Now assume that (\ref{IS}) holds for some $j \in \{1,\ldots,N-1\}$. Upon adding
$[a_{t(j)},b_{t(j)}]$ to $\mathcal{D}(\alp)$, $f$ equals $b_{t(j)}$ in step 3 of the algorithm. Hence
\begin{align*}
f = b_{t(j)} &\leq b_{s(j)}  \qquad \mbox{ since (\ref{IS}) holds for $j$}\\
&< a_{s(j+1)} \leq b_{s(j+1)} \qquad \mbox{ by (\ref{algo4})}
\end{align*}
Since the $b_i$ are non-decreasing, this implies $t(j) <s(j+1)$.  Hence the loop in step 3 will
examine $[a_{s(j+1)},b_{s(j+1)}]$ for possible addition to $ \mathcal{D}(\alp)$ {\sl after} adding
$[a_{t(j)},b_{t(j)}]$. Since $a_{s(j+1)}>f$, the interval $[a_{s(j+1)},b_{s(j+1)}]$ will be added
to $ \mathcal{D}(\alp)$  unless $f$ was changed by adding a different interval
in the meantime (i.e. at some iteration $i \in \{t(j)+1,\ldots,s(j+1)-1\}$ in the for-loop). In either
case an interval $[a_{t(j+1)},b_{t(j+1)}]$ is added to $ \mathcal{D}(\alp)$. Moreover, in either case
$b_{t(j+1)} \leq b_{s(j+1)}$ since the $b_i$ are non-decreasing. Thus (\ref{IS}) holds and therefore
$\# \mathcal{D}(\alp) \geq N(\alp)$. Now the claim about $\mathcal{D}(\alp)$ follows together with
(\ref{algo1}) and (\ref{algo2}). 

Furthermore, every interval $[a_i,b_i] \in \mathcal{D}(\alp)$ must be minimal:
Suppose to the contrary that there exists $[a_k,b_k] \subset \mathcal{C}(\alp)$
such that $[a_k,b_k] \subsetneq [a_i,b_i]$. Due to the sorting of intervals in $\mathcal{C}(\alp)$ we must then
have $k<i$. Hence by (\ref{algo3}), $[a_i,b_i]$ is not the first interval added to $\mathcal{D}(\alp)$, so there exists
an interval $[a_j,b_j] \in \mathcal{D}(\alp)$ that is the last interval added to $\mathcal{D}(\alp)$ before $[a_i,b_i]$.
Hence
\be  \label{algostar}
b_j =: f < a_i
\ee
If $j<k$, then (\ref{algostar}) and $[a_k,b_k] \subsetneq [a_i,b_i]$ give $f=b_j <a_i \leq a_k$, so $[a_k,b_k]$ should
have been added to $\mathcal{D}(\alp)$ after $[a_j,b_j]$ and before $[a_i,b_i]$, which is a contradiction.

If $j\geq k$, then $b_j \geq b_k$ and so (\ref{algostar}) gives $a_i>f=b_j \geq b_k \geq a_k$, hence
$[a_k,b_k]$ cannot be a subset of $[a_i,b_i]$, which is again a contradiction.

In either case, the assumption that $[a_i,b_i] \in \mathcal{D}(\alp)$ is not minimal leads to a contradiction.
\smallskip

It follows immediately from step 3 of the algorithm that all intervals in $\mathcal{D}(\alp)$ are
distinct and that the same holds for $\mathcal{M}(\alp)$.

\section{Proofs}
						
\subsection{Proof of Theorem~\ref{thm:varKnownFinSample}}

The first claim follows since Bonferroni's inequality and (\ref{nulldist}) yield
\begin{align*}
\Pr\Bl( LBD \mbox{ erroneously declares a changepoint } \Br) &=
\Pr \left( \bigcup_{\substack{ \t=(s,m,e) \in \bigcup_{\ell} \KK_{\ell}:\\ \mbox{no CP in $(s,e)$}}}
 \Bl\{ \Tt > c_{\t,n}(\alpha_{\t}) \Br\}\right) \\
&\leq \sum_{\substack{ \t=(s,m,e) \in \bigcup_{\ell} \KK_{\ell}:\\ \mbox{no CP in $(s,e)$}} }
\Pr \Bl(  \Tt > c_{\t,n}(\alpha_{\t}) \Br) \\
&\leq \sum_{\t=(s,m,e) \in \bigcup_{\ell} \KK_{\ell}} \Pr_{\mbox{no CP in $(s,e)$}}\Bl(
\Tt > c_{\t,n}(\alpha_{\t}) \Br) \\
&\leq  \sum_{B=1}^{B_{max}} \sum_{\t \in \mbox{$B$th block}} \dfrac{\alpha}{B \,(\sum_{b=1}^{B_{max}} b^{-1})\
\# (B \text{th block})} \ = \  \alpha
\end{align*}
The second claim follows from the first, as explained in Section~\ref{subsec:varKnown}. $ \hfill \Box$				

\subsection{Proof of Proposition~\ref{complexity}}

Counting the number of triplets in $\mathcal{K}_{\ell}$ can be done similarly to counting the number of Bonferroni intervals
in $\mathcal{J}_{\ell}$, see \cite{walther2021calibrating}[(B7)]:
Since there are no more than $\frac{n}{d_{\ell}}$  left
endpoints for intervals in $\mathcal{J}_{\ell}$ and for each left endpoint there are no more than $\frac{2^{\ell}}{d_{\ell}}$
right endpoints, it follows that $\# \mathcal{J}_{\ell} \leq \frac{n2^{\ell}}{d_{\ell}^2} \leq \frac{2n}{2^{\ell}} \log (en)$
and there are no more than $\frac{2^{\ell}}{d_{\ell}}$ different lengths in $\mathcal{J}_{\ell}$. Thus
\be  \label{cardLn}
\# \mathcal{L}_n\ \leq \  \sum_{\ell=0}^{\lfloor \log_2 \frac{n}{4} \rfloor -1} \frac{2^{\ell}}{d_{\ell}}
\ \leq \ \sum_{\ell=0}^{\log_2 \frac{n}{4}} \sqrt{2 \log \frac{en}{2^{\ell}}}
\ \leq \ 3 \log^{\frac{3}{2}} n
\ee
and therefore
$$
\# \bigcup_{\ell} \mathcal{K}_{\ell}\ \leq \  \sum_{\ell=0}^{\lfloor \log_2 \frac{n}{4} \rfloor -1} 2(\# \mathcal{J}_{\ell})
(\# \mathcal{L}_n)\ \leq \ \sum_{\ell=0}^{ \log_2 \frac{n}{4}} 12 \frac{n \log^{\frac{5}{2}} (en)}{2^{\ell}}
\ \leq \ 24\,n \log^{\frac{5}{2}} (en)
$$

In the exponential family case, (\ref{logLR}) shows that $\log \text{LR}_{\t}(\Yn)$ depends on $\Yn$ only via the partial
sums $\sum_{i=s+1}^m Y_i$, $\sum_{i=m+1}^e Y_i$, and $\sum_{i=s+1}^e Y_i$. (In more detail, the first sup
in (\ref{logLR}) can be computed via the MLE $\hat{\theta}=(A^{\prime})^{-1} \left(\overline{Y}_{(s,m]}\right)$
if the MLE exists, i.e. if $\hat{\theta}$ is finite, and via the Legendre-Fenchel conjugate otherwise.)
Thus $\Tt$ can be computed in $O(1)$ steps if the
vector of cumulative sums $\left( \sum_{i=1}^j Y_i,\, j=1,\ldots,n\right)$ is available. Checking $\Tt>c_{\t,n}
(\alpha)$ and adding $[s+1,e-1]$ to $\mathcal{C}(\alp)$ also requires only $O(1)$ steps. Computing the vector of cumulative sums
once in the beginning requires $O(n)$ steps, and looping over the $O(n \log^{\frac{5}{2}} n)$ Bonferroni triplets $\t$
in order to compute the corresponding $\Tt$  thus results in a computational complexity of 
$O(n \log^{\frac{5}{2}} n)$ for LBD. The same complexity obtains if $\Tt$ is the two-sample $t$-statistic (\ref{two-t}),
as the pooled sample variance can be computed in $O(1)$ steps from the cumulative sums of the $Y_i$ and the
cumulative sums of the $Y_i^2$.

In the case of the Wilcoxon rank-sum statistic it is necessary to compute the local ranks of the $(Y_i,\, i\in I)$
for the various intervals $I=(s,e]$.
One way to do this from scratch, i.e. without speed-up by sharing information across $I$, is to first sort the
$(Y_i,\, i\in I)$ and then in turn insert each $Y_i$ into the sorted list in order to find its rank. Sorting can be
done in $O(|I| \log |I|)$ steps and inserting one $Y_i$ can be done in $O(\log |I|)$ steps via bisection.
Therefore at most $O(|I| \log |I|)$ steps are required for computing the local ranks, followed by $O(|I|)$ steps for computing
the rank sum statistic. Using $|I|\leq n$ and the above bound for $\# \bigcup_{\ell} \mathcal{K}_{\ell}$ shows
that the computational complexity is not worse than $O\left( n^2 \log^{\frac{7}{2}} n\right)$. In fact, a more detailed
accounting for the various sizes of the $|I|$ does not improve this bound. The computational burden can be reduced
by restricting the analysis to intervals with lengths $|I| \leq n^p$ for some $p \in (0,1)$, which seems appropriate
for many practical applications. The resulting computational complexity is then $O(n^{1+p})$, up to log terms.
$ \hfill \Box$

\subsection{Proof of Theorem~\ref{thm:detectgeneral}}

We first give three auxiliary results as lemmata.
Lemma~\ref{approx}(a) is a companion result to \cite{walther2021calibrating}[Prop.~1]
that shows that an interval $I$ can be approximated with a certain precision by a Bonferroni interval $J$
that also satisfies $J \subset I$.

\begin{lem} \label{approx} \noindent
\begin{enumerate}
\item[(a)]
Let $I \subset (0,n]$ be an interval with $1\leq |I| \leq \frac{n}{8}$.
Then we can find a Bonferroni interval $J \subset I$ such that
$$
\frac{ \bigl| J \triangle \,I \bigr|}{|I|}  \leq  \frac{8}{\sqrt{ 2\log \frac{en}{|I|}}}
$$
\item[(b)]
Let $1\leq M \leq  \frac{n}{8} $. Then there exists $L \in \mathcal{L}_n$ such that
$$
 0 \ \leq \ \frac{M-L}{M} \ \leq \ \frac{4}{\sqrt{ 2\log \frac{en}{M}}}
$$
\end{enumerate}
\end{lem}

{\bf Proof of Lemma~\ref{approx}:} For (a), let the integer $\ell$ be defined by $2^{\ell} \leq |I| <2^{\ell +1}$.
Then $0 \leq \ell \leq \lfloor \log_2 \frac{n}{4} \rfloor -1$, so the collection $\mathcal{J}_{\ell}$ exists for this $n$.
If $d_{\ell}=1$, then $I$ itself is a Bonferroni interval and the claim clearly holds. So we assume from
now on that $d_{\ell} >1$. 
Let $j d_{\ell}$ and $k d_{\ell}$ be the smallest resp. largest multiples of $d_{\ell}$ such that
$(j d_{\ell}, k d_{\ell}] \subset I$. 
If $(k-j)d_{\ell} \geq 2^{\ell}$, then $J:= (j d_{\ell}, k d_{\ell}] \in \mathcal{J}_{\ell}$ is a Bonferroni
interval with $|I \triangle J| \leq 2 d_{\ell}$.

If $(k-j)d_{\ell} < 2^{\ell}$, then we employ $\mathcal{J}_{\ell-1}$ in order to find an approximation
(note that $d_{\ell} >1$ implies $\ell \geq 1$): 
Let $\tilde{j} d_{\ell-1}$ and $\tilde{k} d_{\ell-1}$ be the smallest resp. largest multiples of $d_{\ell-1}$ such that
$J:=(\tilde{j} d_{\ell-1}, \tilde{k} d_{\ell-1}] \subset (j d_{\ell}, k d_{\ell}]$. Then $J \in \mathcal{J}_{\ell-1}$
is a Bonferroni interval satsifying $|I \triangle J| \leq 2 d_{\ell-1} +2 d_{\ell} \leq 4  d_{\ell}$. Thus in either case
$$
\frac{ \bigl| J \triangle \,I \bigr|}{|I|}\  \leq \ \frac{4 d_{\ell}}{2^{\ell}} \ \leq \
\frac{8}{\sqrt{2 \log \frac{en}{2^{\ell}}}} \ \leq \ \frac{8}{\sqrt{2 \log \frac{en}{|I|}}}
$$
where we used $\frac{d_{\ell}}{2^{\ell}} \leq \frac{2}{\sqrt{2 \log \frac{en}{2^{\ell}}}}$ for $d_{\ell} >1$,
see the proof of Proposition~1 in \cite{walther2021calibrating}.
\smallskip

For part (b), define the integer $\ell$ by $2^{\ell} \leq M <2^{\ell +1}$, so 
$0 \leq \ell \leq \lfloor \log_2 \frac{n}{4} \rfloor -1$ and thus the collection $\mathcal{J}_{\ell}$ exists for this $n$.
Now the proof proceeds analogously as in part (a): If $d_{\ell}=1$, then we can take $L:=M$. If $d_{\ell} >1$, then
let $k$ be the largest integer such that $k d_{\ell} \leq M$. If $k d_{\ell} \geq 2^{\ell}$, then $L:=k d_{\ell}
\in \mathcal{L}_n$ and $M-d_{\ell} \leq L \leq M$. If $k d_{\ell} < 2^{\ell}$, then we can find $\tilde{k} d_{\ell-1}
\in \mathcal{L}_n$ such that $M -2 d_{\ell} \leq L \leq M$. The claim follows as in (a). $\hfill \Box$

\begin{lem} \label{lem:testStat}
Let $s,e,m,\tau$ be integers with $m \in (s,e)$ and $\tau \in [s,e]$ and let $Y_{s+1}, \ldots, Y_{\tau} \iid N(\mu_1, \sigma^2)$
and $Y_{\tau+1}, \ldots, Y_{e} \iid N(\mu_2, \sigma^2)$.

Set $Q(\mathbf{Y}) := \dfrac{\overline{Y}_{(s,m]}-\overline{Y}_{(m,e]}}{\sigma}\sqrt{\frac{(m-s)(e-m)}{e-s}}$. 
Then $Q(\mathbf{Y}) =Q(\mathbf{\mu}) +Q(\mathbf{Z})$ where $Q(\mathbf{Z}) \sim N(0,1)$ and
\begin{align*}
Q(\mathbf{\mu}) &= \begin{cases}
\dfrac{\mu_1-\mu_2}{\sigma}(e-\tau)\sqrt{\dfrac{(m-s)}{(e-m)(e-s)}}, & m \le \tau \\ 
\dfrac{\mu_1-\mu_2}{\sigma}(\tau-s)\sqrt{\dfrac{(e-m)}{(m-s)(e-s)}}, & \tau < m
\end{cases}
\end{align*} 

\end{lem}

{\bf Proof of Lemma~\ref{lem:testStat}:}
Writing $Y_i = \mu_1 +\sigma Z_i$ for $i \leq \tau$ and $Y_i = \mu_2 +\sigma Z_i$ for $i> \tau$
where $Z_i \iid N(0,1)$, we obtain the stated identity. Straightforward calculations show 
$Q(\mathbf{Z}) \sim N(0,1)$ as well as the expression for $Q(\mathbf{\mu})$. $\hfill \Box$

\begin{lem} \label{critval}
The critical value $z\left(\frac{\alpha_{\t}}{2}\right)$ for the Gaussian case (\ref{eq:critVal}) satisfies
$$
\sup_n \max_{\substack{t=(s,m,e) \in \bigcup_{\ell}\mathcal{K}_{\ell}:\\ (m-s)\wedge (e-m) \leq n^p}}
\left( z\left(\frac{\alpha_{\t}}{2}\right) - \sqrt{2 \log \frac{n}{(m-s)\wedge (e-m)}} \right) \ \leq \ b
$$
for some constant $b$.
\end{lem}

{\bf Proof of Lemma~\ref{critval}:}
This follows from the corresponding result for Bonferroni intervals shown in the proof of Theorem~2 in
\cite{walther2021calibrating}: By the construction of Bonferroni triplets we have $\# \mathcal{K}_{\ell}
\leq 2 (\# \mathcal{J}_{\ell}) (\# \mathcal{L}_n)$. Since $\# \mathcal{L}_n \leq 3 \log^{\frac{3}{2}}n$
by (\ref{cardLn}), it follows that the cardinality of the $B$th block increases at most by a factor of $6 \log^{\frac{3}{2}}n$
as compared to the case of Bonferroni intervals. The proof in \cite{walther2021calibrating} is readily
seen to go through with this additional log factor. $\hfill \Box$
\medskip

For the proof of Theorem~\ref{thm:detectgeneral}, let $\tau_{k_i}$ be a changepoint that satisfies (\ref{eq:st}).
Set $a_i:=\max ( \tau_{k_i}-\del_i, \tau_{k_i-1})$ and $z_i:=\min (\tau_{k_i}+\del_i,\tau_{k_i+1})$, where $\del_i$
is the smallest nonnegative integer such that
\be \label{11}      
|\mu_{\tau_{k_i}+1} -\mu_{\tau_{k_i}}| \, \sqrt{\frac{(\tau_{k_i}-a_i)(z_i-\tau_{k_i})}{z_i-a_i}}
\ \geq \  \sqrt{2\log \frac{n}{(\tau_{k_i}-a_i) \wedge (z_i-\tau_{k_i})}} + \sqrt{2\log m_n} +b_n
\ee
(This definition
amounts to finding the smallest $\delta$ for which the energy between $\tilde{\tau}_{k_i-1}:=\tau_{k_i} -\delta$ and
$\tilde{\tau}_{k_i+1}:=\tau_{k_i} +\delta$ satifies (\ref{eq:st}), with the proviso that $\tilde{\tau}_{k_i-1}$
will not drop below  $\tau_{k_i-1}$ as $\delta$ varies,  and $\tilde{\tau}_{k_i+1}$  will not exceed $\tau_{k_i+1}$.)

This $\del_i$ exists and is not larger than $\max(\tau_{k_i+1} -\tau_{k_i}, \tau_{k_i}-\tau_{k_i-1})$, because plugging in
the latter value for $\del_i$ yields $a_i=\tau_{k_i-1}$ and $ z_i=\tau_{k_i+1}$, for which (\ref{11}) holds because (\ref{eq:st})
hols for $\tau_{k_i}$. 
Further $\del_i \geq 1$ by inspection.

We may assume $\tau_{k_i}-a_i \leq z_i-\tau_{k_i}$, the complementary case being analogous.
Then Lemma~\ref{approx} yields a Bonferroni interval $(s_i,m_i] \subset (a_i,\tau_{k_i}]$ with\footnote{In keeping with earlier notation
we use $m_i$ to denote the middle value of the $i$th Bonferroni triplet, and we alert the reader to keep this separate from $m_n$,
which denotes the number of changepoints we wish
to detect.}
$$
\frac{(s_i-a_i)+(\tau_{k_i}-m_i)}{\tau_{k_i}-a_i}\ \leq \ \frac{8}{\sqrt{2 \log \frac{en}{\tau_{k_i}-a_i}}}\ \leq \ 
\frac{8}{\sqrt{2 (1-p)\log n}}
$$
since $\tau_{k_i}- \tau_{k_i-1} \leq n^p$. Hence
\be  \label{opt1}
\frac{\tau_{k_i}-m_i}{z_i-m_i}\ \leq \ \frac{\tau_{k_i}-m_i}{\tau_{k_i}-a_i} \ \leq \  \frac{8}{\sqrt{2 (1-p)\log n}}
\ee
as $z_i-m_i \geq z_i-\tau_{k_i} \geq \tau_{k_i}-a_i$. Furthermore
\be  \label{opt15}
\frac{m_i-s_i}{\tau_{k_i}-a_i}\ =\ 1- \frac{\tau_{k_i}-a_i -m_i +s_i}{\tau_{k_i}-a_i}\ \geq \ 1-\frac{8}{\sqrt{2 (1-p)\log n}}
\ee
Part (b) of the Lemma yields an integer $e_i \in (m_i,z_i]$ such that $e_i-m_i \in \mathcal{L}_n$ and
\be  \label{opt2}
\frac{z_i-e_i}{z_i-m_i}\ =\ \frac{(z_i-m_i) -(e_i-m_i)}{z_i-m_i}\ \leq \ \frac{4}{\sqrt{2 \log \frac{en}{z_i-m_i}}}
\ \leq \ \frac{4}{\sqrt{2 (1-p)\log n}}
\ee
Thus $(s_i,m_i,e_i)$ is a Bonferroni triplet ($m_i-s_i \leq e_i-m_i$ can be shown to follow from $\tau_{k_i}-a_i \leq z_i-\tau_{k_i}$
and (\ref{opt2}), and by reducing $m_i$ if necessary.) (\ref{opt1}) and (\ref{opt2}) yield
$$
e_i -\tau_{k_i} = (e_i-z_i)+(z_i-m_i)+(m_i-\tau_{k_i}) \geq \left(1-\frac{4+8}{\sqrt{2 (1-p)\log n}}\right) (z_i-m_i) >0
$$
for $n$ large enough (not depending on $\tau_{k_i}$). Since $s_i<\tau_{k_i}$ by construction, it follows that the
changepoint $\tau_{k_i}$ falls into $(s_i,e_i)$. Conversely, there is no further changepoint in $(s_i,e_i)$
 since $\tau_{k_i-1}$ is not larger than $a_i\leq s_i$, 
while $\tau_{k_i+1}$ is not smaller than $z_i$ and $z_i\geq e_i$. Therefore the assumption of
Lemma~\ref{lem:testStat} are met and the test statistic $T_{s_i m_i e_i}(\Yn)$ in (\ref{localtest}) satisfies
$$
T_{s_i m_i e_i}(\Yn) \ \geq \ |\mu_{\tau_{k_i}+1} -\mu_{\tau_{k_i}}| \, (e_i-\tau_{k_i})\sqrt{\frac{m_i-s_i}{(e_i-m_i)(e_i-s_i)}}
-|N(0,1)|
$$
Now (\ref{opt2}) and $z_i-m_i \leq z_i-a_i \leq 2(z_i-\tau_{k_i})$ yield
$$
e_i-\tau_{k_i} \geq z_i-\tau_{k_i} -\frac{4}{\sqrt{2 (1-p)\log n}} (z_i-m_i) \geq \left(1-\frac{8}{\sqrt{2 (1-p)\log n}}\right)
(z_i-\tau_{k_i})
$$
while (\ref{opt1}) and $z_i-m_i \leq 2(z_i-\tau_{k_i})$ give
$$
e_i-m_i \leq (z_i-\tau_{k_i}) +(\tau_{k_i}-m_i) \leq \left(1+\frac{16}{\sqrt{2 (1-p)\log n}}\right) (z_i-\tau_{k_i})
$$
Together with (\ref{opt15}), $e_i-s_i \leq z_i-a_i$ and (\ref{11}) we obtain
\begin{align*}
T_{s_i m_i e_i}(\Yn) \ &\geq \ |\mu_{\tau_{k_i}+1} -\mu_{\tau_{k_i}}| \, \sqrt{\frac{(z_i-\tau_{k_i})(\tau_{k_i}-a_i)}{z_i-a_i}}
\frac{\left(1-\frac{8}{\sqrt{2 (1-p)\log n}}\right)^{\frac{3}{2}}}{\sqrt{1+\frac{16}{\sqrt{2 (1-p)\log n}}}} -|N(0,1)|\\
&\geq \left(\sqrt{2\log \frac{n}{(\tau_{k_i}-a_i) \wedge (z_i-\tau_{k_i})}} + \sqrt{2\log m_n} +b_n \right) 
\left(1-\frac{24}{\sqrt{2 (1-p)\log n}}\right) -|N(0,1)|\\
&\geq \sqrt{2\log \frac{n}{(\tau_{k_i}-a_i) \wedge (z_i-\tau_{k_i})}} + \sqrt{2\log m_n} +\frac{b_n}{2} -|N(0,1)|
\end{align*}

for $n$ large enough (depending only on $b_n$), since $m_n \leq n$ implies
$$
24 \frac{\sqrt{2\log \frac{n}{(\tau_{k_i}-a_i) \wedge (z_i-\tau_{k_i})}} + \sqrt{2\log m_n}}{\sqrt{2 (1-p)\log n}} 
\leq \frac{48}{\sqrt{1-p}}
$$
Now $\frac{|m_i-s_i|}{\tau_{k_i}-a_i} \in (\frac{1}{2},1]$ by (\ref{opt15}), hence Lemma~\ref{critval} gives
\be  \label{opt4}
z\left(\frac{\alpha_{s_i m_i e_i} }{2}\right) \leq \sqrt{2\log \frac{n}{(\tau_{k_i}-a_i) \wedge (z_i-\tau_{k_i})}} +\gamma
\ee
for some constant $\gamma$, uniformly in $n$ and in $i=1,\ldots, m_n$.

If $T_{s_i m_i e_i}(\Yn) \geq z\left(\frac{\alpha_{s_i m_i e_i} }{2}\right)$, then LBD gives the confidence
interval $[s_i+1,e_i-1]$ for a changepoint. Since it was shown above that $\tau_{k_i} \in [s_i+1,e_i-1]$, LBD
thus localizes $\tau_{k_i}$ with precision not worse than $\max\left( (e_i-1)-\tau_{k_i}, \tau_{k_i} -(s_i+1)\right)
\leq \max \left(z_i-\tau_{k_i}-1,\tau_{k_i}-a_i-1\right) \leq \del_i-1$ by the definition of $a_i$ and $z_i$ above.
Therefore
\begin{align*}      
\Pr_{\bs{\mu}_n} &\Bigl(\mbox{LBD detects all $\tau_{k_1},\ldots,\tau_{k_{m_n}}$
                and localizes each $\tau_{k_i}$  with precision $\del_i-1$}\Bigr) \\
&\geq 1- \Pr_{\bs{\mu}_n} \left( \bigcup_{i=1}^{m_n} \Bigl\{ T_{s_i m_i e_i}(\Yn) < 
z\left(\frac{\alpha_{s_i m_i e_i} }{2}\right) \Bigr\} \right)
\end{align*}
and the Gaussian tail bound gives
\begin{align*}
\Pr_{\bs{\mu}_n} &\left( \bigcup_{i=1}^{m_n} \Bigl\{ T_{s_i m_i e_i}(\Yn) <z\left(\frac{\alpha_{s_i m_i e_i} }{2}\right)
 \Bigr\} \right) \\
&\leq   \sum_{i=1}^{m_n} \Pr \Bigl( |N(0,1)| > \sqrt{2\log \frac{n}{(\tau_{k_i}-a_i) \wedge (z_i-\tau_{k_i})}} + \sqrt{2\log m_n} 
+\frac{b_n}{2} -z\left(\frac{\alpha_{s_i m_i e_i} }{2}\right)\Bigr) \\
&\leq \sum_{i=1}^{m_n} \Pr \left( |N(0,1)| > \sqrt{2\log m_n} +\frac{b_n}{2} -\gamma \right) \qquad \mbox{ by (\ref{opt4})}\\
&\leq \sum_{i=1}^{m_n} 2\exp \Bigl\{ -\frac{1}{2} \Bigl(\sqrt{2 \log m_n} +\frac{b_n}{2} -\gamma \Bigr)^2 \Bigr\} \\
&= 2\exp \Bigl\{ -\frac{1}{2} \Bigl(\frac{b_n}{2} -\gamma \Bigr)^2 - \sqrt{2 \log m_n} 
  \Bigl(\frac{b_n}{2} -\gamma\Bigr) \Bigr\} \ \rightarrow 0 
\end{align*}

In order to establish (\ref{precision}) we define for real $\del >0$: $a_i(\del):=\max ( \tau_{k_i}-\del,\tau_{k_i-1}),
z_i(\del):=\min (\tau_{k_i}+\del,\tau_{k_i+1})$. Without loss of generality we may assume $\mm_i=
\tau_{k_i}-\tau_{k_i-1} \leq \tau_{k_i+1} -\tau_{k_i}$. Then we consider the difference of the two sides of (\ref{11}):
\begin{align}
F_i(\del):= |\mu_{\tau_{k_i}+1} -\mu_{\tau_{k_i}}| \, \sqrt{\frac{(\tau_{k_i}-a_i(\del))(z_i(\del)-\tau_{k_i})}{z_i(\del)-a_i(\del)}}
- \sqrt{2\log \frac{n}{(\tau_{k_i}-a_i(\del)) \wedge (z_i(\del)-\tau_{k_i})}} - \sqrt{2\log m_n} -b_n  \nn \\
= \begin{cases} |\mu_{\tau_{k_i}+1} -\mu_{\tau_{k_i}}| \sqrt{\frac{\del}{2}} -\sqrt{ 2 \log \frac{n}{\del}} - \sqrt{2\log m_n} -b_n
& \text{ if $\del \in (0,\tau_{k_i}-\tau_{k_i-1}]$} \\
|\mu_{\tau_{k_i}+1} -\mu_{\tau_{k_i}}| \sqrt{\frac{(\tau_{k_i}-\tau_{k_i-1})\, \del}{\tau_{k_i}-\tau_{k_i-1} +\del}}
-\sqrt{ 2 \log \frac{n}{\tau_{k_i}-\tau_{k_i-1}}} - \sqrt{2\log m_n} -b_n
& \text{ if $\del \in (\tau_{k_i}-\tau_{k_i-1},\tau_{k_i+1}-\tau_{k_i}]$} 
\end{cases}
\label{1st}
\end{align}

Since the changepoint $\tau_{k_i}$ satisfies (\ref{eq:st}), we have $|\mu_{\tau_{k_i}+1} -\mu_{\tau_{k_i}}|>0$ and
$F_i(\tau_{k_i+1}-\tau_{k_i}) \geq 0$. Thus $F_i(\cdot)$ is continuous and strictly increasing for $\del \in (0,
\tau_{k_i+1}-\tau_{k_i}]$, and $F_i(\del)<0$ for $\del$ near 0. Therefore
$\del_i^* := \inf \left\{ \del >0: \ F_i(\del) \geq 0 \right\}$
satifies $\del_i^* \in (0,\tau_{k_i+1}-\tau_{k_i}]$ and $F_i(\del_i^*)=0$ and
\be \label{2st}
 \del_i -1 \ \leq \ \del_i^*
\ee
since $\del_i$ was defined as the smallest {\sl integer} satisfying $F_i(\del_i) \geq 0$.

The condition in the first case of (\ref{precision}) says that $F_i(\tau_{k_i}-\tau_{k_i-1}) \geq 0$.
Therefore $\del_i^* \leq \tau_{k_i}-\tau_{k_i-1}$ and so $F_i(\del_i^*)=0$ and (\ref{1st}) give
$$
\del_i^* \ =\ 2 \left( \frac{ \sqrt{ 2 \log \frac{n}{\del_i^*}} + \sqrt{2\log m_n} +b_n}{|\mu_{\tau_{k_i}+1} -\mu_{\tau_{k_i}}|}
\right)^2\ \geq \ |\mu_{\tau_{k_i}+1} -\mu_{\tau_{k_i}}|^{-2},
$$
hence $\del_i^* =2g(\del_i^* ) \leq 2 g\left( |\mu_{\tau_{k_i}+1} -\mu_{\tau_{k_i}}|^{-2}\right)$ since the function $g$
is decreasing.

If the condition in the first case of (\ref{precision}) does not hold, 
then $F_i(\tau_{k_i}-\tau_{k_i-1}) <0$ and hence $\del_i^* > \tau_{k_i}-\tau_{k_i-1}$.
Then $F_i(\del_i^*)=0$ and (\ref{1st}) give
$$
\frac{(\tau_{k_i}-\tau_{k_i-1})\, \del_i^*}{\tau_{k_i}-\tau_{k_i-1} +\del_i^*}\ =\ g(\tau_{k_i}-\tau_{k_i-1})
$$
which yields
$$
\del_i^* \ = \ \left( 1-\frac{g(\tau_{k_i}-\tau_{k_i-1})}{\tau_{k_i}-\tau_{k_i-1}}\right)^{-1} g(\tau_{k_i}-\tau_{k_i-1})
$$
(\ref{precision}) now follows with (\ref{2st}).
$\hfill \Box$

\subsection{Proof of Theorem \ref{thm:lb-single}}
We will use the following elementary property of a finite sequence:
\begin{lem} \label{L1}
Let $v_n \in \bf{N}$ be such that $v_n \leq \frac{n}{4}$ and set $\del_n:= \lfloor \frac{n}{4v_n}\rfloor$.
If $\bs{\mu}_n$ has not more than $v_n$ changepoints, then there exist $v_n$ integers
$t_i$, $i=1,\ldots,v_n$, such 
that the sets $I_i:=\{t_i -\del_n+1,\ldots, t_i +\del_n \} \subset \{1,\ldots,n\}$ are pairwise disjoint
and $\bs{\mu}_n$ is constant on each $I_i$. 
\end{lem}
							
A salient consequence of Lemma~\ref{L1} is the fact that $\sum_{i=1}^{v_n} \# I_i$ is order $n$.
This is key for applying the weak law of large numbers below.
						
{\bf Proof of Lemma~\ref{L1}:} Let the changepoints of $\bs{\mu}_n$ be 
$\tau_1, \ldots, \tau_{K_n}$ with $0 =: \tau_0 < \tau_1 < \ldots < \tau_{K_n} < \tau_{K_n+1} := n$.
Then we can find $v_n$ pairwise disjoint $I_i$ with the required length if 
$\sum_{i=1}^{K_n+1} \lfloor \frac{\tau_i-\tau_{i-1}}{2\delta_n} \rfloor \ge v_n$ holds. Now
$$
\sum_{i=1}^{K_n+1} \left\lfloor \frac{\tau_i-\tau_{i-1}}{2\delta_n} \right\rfloor \ >\ \sum_{i=1}^{K_n+1} 
\left( \frac{\tau_i-\tau_{i-1}}{2\delta_n} -1 \right)\ =\ \dfrac{n}{2\delta_n}-(K_n+1) \ \ge \ \dfrac{n}{2\delta_n}-(v_n+1).
$$
Since $\frac{n}{4\delta_n} \ge v_n$  we obtain
$\dfrac{n}{2\delta_n}-(v_n+1) \ge v_n  - 1$. Since $\sum_{i=1}^{K_n+1} \lfloor
\ldots \rfloor $ is an integer, it must therefore be at least as large as $v_n$. $ \hfill \Box$
\bigskip
							
Let $t_i$ and $I_i$, $i=1,\ldots,v_n$, and $\del_n$ as given by Lemma~\ref{L1}. For $i=1,\ldots,v_n$ set
$$
\bs{\mu}_n^{(i)} := \bs{\mu}_n +\frac{1}{2} \sqrt{ \frac{(2-\eps_n)^2}{\del_n} \log \frac{n}{\del_n}}\
\Bl( \bs{1}_{\{t_i +1,\ldots,t_i +\del_n\}} - \bs{1}_{\{t_i -\del_n+1,\ldots,t_i \}} \Br)
$$
where $\{\eps_n\}$ is any sequence with $\eps_n \ra 0$ and $\eps_n \sqrt{\log \frac{n}{\del_n}} \ra \infty$.
Then the changepoints of $\bs{\mu}_n$ are also changepoints of $\bs{\mu}_n^{(i)}$, and $\bs{\mu}_n^{(i)}$ has changepoints at 
$\tau_{k_i}^{(i)}:=t_i$ and at $t_i -\del_n$ and at $t_i+\del_n$.
Since $\bs{\mu}_n$ is constant on $I_i$ it follows that $\tau_{k_i}^{(i)}$ is {\sl not} a changepoint of $\bs{\mu}_n$
and that the distance from $\tau_{k_i}^{(i)}$
to its preceding and succeeding changepoints is $\del_n$. Hence
$$
\sqrt{\frac{\left(\tau_{k_i}^{(i)}-\tau_{k_i-1}^{(i)}\right)\left(\tau_{k_i+1}^{(i)}-
\tau_{k_i}^{(i)}\right)}{\tau_{k_i+1}^{(i)}-\tau_{k_i-1}^{(i)}}} = \sqrt{\frac{\del_n}{2}}
$$
and therefore
\begin{align*}
\left|\mu_{\tau_{k_i}^{(i)} +1}^{(i)} -\mu_{\tau_{k_i}^{(i)}}^{(i)}\right| \,
\sqrt{\frac{\left(\tau_{k_i}^{(i)}-\tau_{k_i-1}^{(i)}\right)\left(\tau_{k_i+1}^{(i)}-
\tau_{k_i}^{(i)}\right)}{\tau_{k_i+1}^{(i)}-\tau_{k_i-1}^{(i)}}} &=  
\sqrt{\frac{(2-\eps_n)^2}{\del_n} \log \frac{n}{\del_n}} \ \sqrt{\frac{\del_n}{2}}\\
&= \left( \sqrt{2} -\frac{\eps_n}{\sqrt{2}} \right)
\sqrt{\log\frac{n}{\left(\tau_{k_i}^{(i)}-\tau_{k_i-1}^{(i)}\right) \wedge
\left(\tau_{k_i+1}^{(i)}-\tau_{k_i}^{(i)}\right)}}
\end{align*}
Thus (\ref{1a}) holds. Write $\phi(\cdot)$ for the standard normal density and
$$
L_{ni}(\bs{Y}_n)\ :=\ \prod_{j=1}^n \frac{\phi(Y_j -\mu_j^{(i)} )}{\phi(Y_j -\mu_j )}
$$
for the likelihood ratio. The following argument is classical, see e.g. \cite{lepski2000asymptotically} and
\cite{dumbgen2001multiscale}: For any test $\psi_n (\bs{Y}_n)$ with $\Ex_{\bs{\mu}_n} \psi_n (\bs{Y}_n)
\leq \alp +o(1)$:
\begin{align*}
\min_{i=1,\ldots,v_n} \Ex_{\bs{\mu}_n^{(i)}} \psi_n (\bs{Y}_n) -\alp &\leq \frac{1}{v_n} \sum_{i=1}^{v_n}
\Ex_{\bs{\mu}_n^{(i)}} \psi_n (\bs{Y}_n) -\Ex_{\bs{\mu}_n} \psi_n (\bs{Y}_n) +o(1)\\
&= \frac{1}{v_n} \sum_{i=1}^{v_n} \Ex_{\bs{\mu}_n} \psi_n (\bs{Y}_n) L_{ni} -\Ex_{\bs{\mu}_n} \psi_n (\bs{Y}_n) +o(1)\\
&= \Ex_{\bs{\mu}_n} \left( \left(\frac{1}{v_n} \sum_{i=1}^{v_n} L_{ni} -1 \right) \psi_n (\bs{Y}_n)\right) +o(1)\\
& \leq \Ex_{\bs{\mu}_n} \left|\frac{1}{v_n} \sum_{i=1}^{v_n} L_{ni} -1 \right| +o(1)
\end{align*}
							
The goal is to show 
\be  \label{1b}
\Ex_{\bs{\mu}_n} \left|\frac{1}{v_n} \sum_{i=1}^{v_n} L_{ni} -1 \right| \ \ra \ 0 \qquad (n \ra \infty)
\ee
Write $\gamma_n := \frac{1}{2}\sqrt{\frac{(2-\eps_n)^2}{\del_n} \log \frac{n}{\del_n}}$. 
Note that $\bs{\mu}_n$ and $\bs{\mu}_n^{(i)}$ coincide except for the coordinates $j \in I_i$, where
$\mu_j^{(i)}=\mu_j -\gamma_n$ for the first $\del_n$ indices $j$ and $\mu_j^{(i)}=\mu_j +\gamma_n$
for the last $\del_n$ indices $j$. Hence
\begin{align*}
L_{ni} &= \prod_{j \in I_i} \exp \left\{ -\frac{1}{2} \left(Y_j -(\mu_j \mp \gamma_n)\right)^2
+\frac{1}{2} \left(Y_j -\mu_j \right)^2 \right\} \\
& \mbox{where the '$-$' sign applies for the first $\del_n$ indices and the '$+$' sign for the
last $\del_n$ indices}\\
&= \exp \left\{ \sum_{j \in I_i} \left( -\frac{1}{2} (Z_j \pm \gamma_n)^2 +\frac{1}{2} Z_j^2 \right) \right\}\\
&= \exp \left\{ \sum_{j \in I_i} \left( \mp \gamma_n Z_j - \frac{1}{2} \gamma_n^2\right) \right\} \\
& = \exp \left\{ \sqrt{2 \del_n \gamma_n^2}\, X_i -\del_n \gamma_n^2 \right\} \qquad \mbox{ since $\# I_i=2\del_n$}
\end{align*}
where $X_i := (2 \del_n \gamma_n^2)^{-\frac{1}{2}} \sum_{j \in I_i} \mp \gamma_n Z_j \sim N(0,1)$.
Note that the $X_i$ are independent since the $I_i$ are pairwise disjoint. Hence (\ref{1b}) follows from
Lemma~6.2 in \cite{dumbgen2001multiscale} upon checking that
\be  \label{condition}
\eta_n:= 1 -\frac{\sqrt{2 \del_n \gamma_n^2}}{\sqrt{2 \log v_n}} \ra 0 \qquad \mbox{ and } \qquad
\sqrt{\log v_n}\, \eta_n \ra \infty \ \ (n \ra \infty)
\ee
since $v_n \ra \infty$\footnote{This is the reason why $v_n$ is required to diverge to infinity.}.  
$\del_n= \lfloor \frac{n}{4v_n}\rfloor \geq 1$ implies that $v_n=r_n \frac{n}{\del_n}$ with $r_n \in [\frac{1}{8},
\frac{1}{4}]$. Therefore
$$
\frac{\del_n \gamma_n^2}{\log v_n}\ =\ \frac{(2-\eps_n)^2 \log \frac{n}{\del_n}}{4 \log \left( r_n \frac{n}{\del_n} 
\right)}\ =\ \left( 1-\eps_n+\frac{\eps_n^2}{4} \right) (1+o(1))\ =\ 1+o(1)
$$
since $\frac{n}{\del_n} \ra \infty$ and $\eps_n \ra 0$, hence the first condition in (\ref{condition}) is met.  
As for the second condition,
\begin{align*}
\sqrt{ \log v_n}\, \eta_n &= \sqrt{ \log v_n} -\sqrt{\del_n \gamma_n^2} \ =\ 
\sqrt{ \log \left( r_n\frac{n}{\del_n} \right)} - \frac{2-\eps_n}{2} \sqrt{\log \frac{n}{\del_n}}\\
&= \frac{\log \left( r_n \frac{n}{\del_n} \right) -\log  \frac{n}{\del_n} }{
\sqrt{\log \left( r_n\frac{n}{\del_n} \right)} +\sqrt{\log \frac{n}{\del_n}}} +
\frac{\eps_n}{2} \sqrt{\log \frac{n}{\del_n}}\\
&= o(1) +\frac{\eps_n}{2} \sqrt{\log \frac{n}{\del_n}} \ \ra \infty \qquad \ \ \mbox{ since } r_n \in \left[\frac{1}{8},
\frac{1}{4}\right]
\end{align*}
Finally we note that in our construction of $\bs{\mu}_n^{(i)}$ the changepoint $\tau_{k_i}^{(i)}$ is equidistant 
from the preceding and
succeeding changepoints. This was done in order to keep the exposition simple, but the proof goes through analogously
with any prescribed ratio of these two distances.
$\hfill \Box$

\subsection{Proof of Theorem \ref{thm:lb-simul}}
We set $\del_n:= \lfloor \frac{n}{4m_n}\rfloor$ so that we
can reuse parts of the proof of Theorem~\ref{thm:lb-single}.
For $i=1,\ldots,m_n$ set $t_i:=(2i-1)\del_n$, $I_i:=\{t_i-\del_n+1,\ldots,t_i+\del_n\}$ and
\be  \label{simulst}
\bs{\mu}_n := \sum_{i=1}^{m_n} \frac{1}{2} \frac{(4-\eps_n)}{\sqrt{\del_n}} \sqrt{\log \frac{n}{\del_n}}\
\Bl( \bs{1}_{\{t_i +1,\ldots,t_i +\del_n\}} - \bs{1}_{\{t_i -\del_n+1,\ldots,t_i \}} \Br)
\ee
where $\eps_n\to0$ satisfies $\eps_n \sqrt{\log \frac{n}{\del_n}} \ra \infty$. Then the changepoints $\tau_{k_i}:=
t_i$ have distance $\del_n$ from their preceding and succeeding changepoints and so (\ref{eq:lb-simulreq}) holds
 as required. (There are additional changepoints
in $\bs{\mu}_n$ but our analysis concerns only inference about the $\tau_{k_i}$, $i=1,\ldots,m_n$.) 
							
Define $\bs{\mu}_n^{(i)}$ in the same way as $\bs{\mu}_n$ but with the jump at $t_i$ only half as large:
$$
\mu_j^{(i)} = \begin{cases} \mu_j & \mbox{if }j \in \{1,\ldots,n\} \setminus I_i \\
\frac{1}{2} \frac{(2-\frac{\eps_n}{2})}{\sqrt{\del_n}} \sqrt{\log \frac{n}{\del_n}}\
\Bl( \bs{1}_{\{t_i +1,\ldots,t_i +\del_n\}} - \bs{1}_{\{t_i -\del_n+1,\ldots,t_i \}} \Br) & \mbox{if }j \in I_i
\end{cases}
$$
Finally, define $\bs{\nu}_n^{(i)}$ as
$$
\nu_j^{(i)} = \begin{cases} \mu_j^{(i)} & \mbox{if }j \in I_i\\
0 & \mbox{if }j \in \{1,\ldots,n\} \setminus I_i  \end{cases}
$$
We will show:
\begin{align} 
\Pr_{\bs{\mu}_n} \Bl( \psi_n (\bs{Y}_n) & \mbox{ detects all changepoints }t_j, j=1,\ldots,m_n \Br) \nn \\
& = \frac{1}{m_n} \sum_{i=1}^{m_n} \Pr_{\bs{\mu}_n^{(i)}} \Bl( \psi_n (\bs{Y}_n) \mbox{
detects all changepoints }t_j, j=1,\ldots,m_n \Br) +o(1) \label{simul1*}
\end{align}
						
\be \label{simul2*}
\Pr_{\bs{\mu}_n^{(i)}} \Bl( \psi_n (\bs{Y}_n) \mbox{ detects }t_i\Br) \ =\ 
\Pr_{\bs{\nu}_n^{(i)}} \Bl( \psi_n (\bs{Y}_n) \mbox{ detects }t_i\Br),\ i=1,\ldots,m_n
\ee
Therefore
\begin{align*}
\Pr_{\bs{\mu}_n} \Bl( \psi_n (\bs{Y}_n) & \mbox{ detects all changepoints }t_j, j=1,\ldots,m_n \Br)  \\
& = \frac{1}{m_n} \sum_{i=1}^{m_n} \Pr_{\bs{\mu}_n^{(i)}} \Bl( \psi_n (\bs{Y}_n) \mbox{
detects all changepoints }t_j, j=1,\ldots,m_n \Br) +o(1)\\
& \leq \frac{1}{m_n} \sum_{i=1}^{m_n} \Pr_{\bs{\mu}_n^{(i)}} \Bl( \psi_n (\bs{Y}_n) \mbox{ detects }t_i\Br)+o(1)\\
& = \frac{1}{m_n} \sum_{i=1}^{m_n} \Pr_{\bs{\nu}_n^{(i)}} \Bl( \psi_n (\bs{Y}_n) \mbox{ detects }t_i\Br)+o(1)\\
& \leq \frac{1}{m_n} \sum_{i=1}^{m_n} \Pr_{\bs{\nu}_n^{(i)}} \Bl( \psi_n (\bs{Y}_n) \mbox{ detects any 
changepoint}\Br)+o(1)\\
& \leq \alpha +o(1)
\end{align*}
where the last inequality follows from the proof of Theorem~\ref{thm:lb-single} with $v_n:=m_n$:
The changepoint $t_i$ in $\bs{\nu}_n^{(i)}$ is exactly as in the proof of Theorem~\ref{thm:lb-single},
and while the statement of Theorem~\ref{thm:lb-single} is about $\min_i$, the proof is
for the average over $i$. It remains to show (\ref{simul1*}) and (\ref{simul2*}).
							
We will prove (\ref{simul1*}) by showing that $\bs{\mu}_n$ and $\frac{1}{m_n} \sum_{i=1}^{m_n} \bs{\mu}_n^{(i)}$
are asymptotically indistinguishable; this proof is almost identical to that of Theorem~\ref{thm:lb-single}:
We calculate the likelihood ratio as
$$
L_{ni}(\bs{Y}_n)\ :=\ \prod_{j=1}^n \frac{\phi(Y_j -\mu_j^{(i)} )}{\phi(Y_j -\mu_j )}\ =\ 
\exp \left\{ \sqrt{2 \del_n \gamma_n^2}\, X_i -\del_n \gamma_n^2 \right\}
$$
where $\gamma_n:= \frac{1}{2} \frac{2-\frac{\eps_n}{2}}{\sqrt{\del_n}} \sqrt{ \log \frac{n}{\del_n}}$
and the $X_i$ are i.i.d. N(0,1). Therefore
$$
\Ex_{\bs{\mu}_n} \left| \frac{1}{m_n} \sum_{i=1}^{m_n} L_{ni} -1 \right| \ra 0 \qquad (n \ra \infty)
$$
follows from the proof of Theorem~\ref{thm:lb-single} with $v_n:=m_n$ since $\gamma_n^2$ equals the value
of $\gamma_n^2$ used there and the choice of $\del_n$ is also the same. Therefore
\begin{align*}
\Bl| & \Pr_{\bs{\mu}_n} \Bl( \psi_n (\bs{Y}_n)  \mbox{ detects all }t_j\Br) - 
\frac{1}{m_n} \sum_{i=1}^{m_n} \Pr_{\bs{\mu}_n^{(i)}} \Bl( \psi_n (\bs{Y}_n) \mbox{ detects all }t_j\Br)\Br|\\
&= \Bl| \Ex_{\bs{\mu}_n} {\bs 1} \Bl( \psi_n (\bs{Y}_n)  \mbox{ detects all }t_j\Br) -
\frac{1}{m_n} \sum_{i=1}^{m_n} \Ex_{\bs{\mu}_n} {\bs 1} \Bl( \psi_n (\bs{Y}_n) \mbox{ detects all }t_j\Br) 
L_{ni}(\bs{Y}_n) \Br|\\
&\leq \Ex_{\bs{\mu}_n} \Bl| 1-\frac{1}{m_n} \sum_{i=1}^{m_n} L_{ni} \Br| \ra 0
\end{align*}
proving (\ref{simul1*}).
							
As for (\ref{simul2*}), note that $\{Y_j, j \in I_i\}$ is a sufficient statistic for detecting $t_i$
and the distribution of $\{Y_j, j \in I_i\}$ does not depend on $\{\mu_j, j \not\in I_i\}$. We may
therefore presume that $\Pr_{\bs{\mu}_n} \Bl( \psi_n (\bs{Y}_n) \mbox{ detects }t_i\Br)$
does not depend on $\{\mu_j, j \not\in I_i\}$, yielding (\ref{simul2*}).
Alternatively, one can establish
$$
\frac{1}{m_n} \sum_{i=1}^{m_n} \Pr_{\bs{\mu}_n^{(i)}} \Bl( \psi_n (\bs{Y}_n) \mbox{ detects }t_i\Br) =\alpha +o(1)
$$
by extending the proof of Theorem~\ref{thm:lb-single} to likelihood ratios with respect to varying
$\widetilde{\bs{\mu}}_n^{(i)}$ given by $\widetilde{\mu_j}^{(i)}=\mu_j$ for $j \in \{1,\ldots,n\}\setminus I_i$
and $\widetilde{\mu_j}^{(i)}=0$ otherwise. $ \hfill \Box$

\subsection{Proof of Theorem \ref{thm:numcpt}}

The proof follows that of Theorem~\ref{thm:detectgeneral}. Recall that this proof proceeds by showing
that if $\tau_{k_i}$ satisfies (\ref{eq:st}), then one can find a shortest interval $(a_i,z_i) \subset
(\tau_{k_i -1},\tau_{k_i+1})$ such that (\ref{11}) holds, and furthermore there exists a Bonferroni triplet
$(s_i,m_i,e_i)$ such that $(s_i,e_i) \subset (a_i,z_i)$ and  $T_{s_i m_i e_i}(\Yn)$ will be significant,
yielding the confidence interval $[s_i +1, e_i -1]$ for $\tau_{k_i}$.

In order to produce confidence intervals that are disjoint, we will cut the widths of these intervals in half:
Define $\tilde{a}_i:= \tau_{k_i} - \left\lceil \frac{\tau_{k_i} -a_i}{2} \right\rceil$ and
$\tilde{z}_i:= \tau_{k_i} + \left\lceil \frac{z_i -\tau_{k_i} }{2} \right\rceil$. Then
$$
\sqrt{\frac{ (\tau_{k_i} -\tilde{a}_i) (\tilde{z}_i -\tau_{k_i})}{\tilde{z}_i -\tilde{a}_i}}
\geq \frac{1}{\sqrt{2}}\sqrt{\frac{ (\tau_{k_i} -a_i) (z_i -\tau_{k_i})}{z_i -a_i}}
$$
So (\ref{11}) continues to hold with $\tilde{a}_i$ and $\tilde{z}_i$ in place of $a_i$ and $z_i$
because (\ref{detectdisjoint}) provides an additional factor $\sqrt{2}$ when compared to (\ref{eq:st}).
The resulting confidence interval for $\tau_{k_i}$ is $[\tilde{s}_i +1,\tilde{e}_i-1] \subset
[\tilde{a}_i +1,\tilde{z}_i-1]$. These confidence intervals are disjoint since $a_{i+1} \geq \tau_{k_i}$
and $z_i \leq \tau_{k_i+1}$ imply
$$
(\tilde{a}_{i+1} +1) - (\tilde{z}_i-1)\ >\ \tau_{k_i+1} -\frac{\tau_{k_i+1} -a_{i+1}}{2} -
\left( \tau_{k_i} +\frac{z_i -\tau_{k_i}}{2} \right)\ =\ \frac{1}{2} (\tau_{k_i+1} -z_i)
+\frac{1}{2} (a_{i+1} -\tau_{k_i}) \ \geq \ 0
$$

Therefore we obtain
$$
\Pr_{\bs{\mu}_n} \Bigl(\mathcal{A}_n:=\left\{\mbox{LBD detects all $\tau_1,\ldots,\tau_{K_n}$
      and produces disjoint confidence intervals for the $\tau_i$ }\right\}\Bigr)
\ \ra \ 1
$$

Since on the event $\mathcal{A}_n$ there are $K_n$ disjoint confidence intervals, 
the definition of $N(\alp)$ gives $N(\alp)\geq K_n$ on $\mathcal{A}_n$.
On the other hand, $\Pr_{\bs{\mu}_n} \Bigl(N(\alp) \leq K_n\Bigr) \geq 1-\alp$ by Theorem~\ref{thm:varKnownFinSample}.
The second claim of the theorem follows. 
$ \hfill \Box$

\section{Details on the test models used in the numerical experiments} \label{a.simul.fun}
We include the details of the signal functions in \cite{fryzlewicz2014wild} for completeness:
\begin{enumerate}
\item blocks: length 2048, changepoints at 205, 267, 308, 472, 512, 820, 902, 1332, 1557, 1598, 1659. Values between changepoints:
 0, 14.64, -3.66, 7.32, -7.32, 10.98, -4.39, 3.29, 19.03 ,7.68, 15.37, 0. Standard deviation of the noise: $\sigma = 10$.
\item fms: length 497, changepoints at 139, 226, 243, 300, 309, 333. Values between changepoints: -0.18, 0.08, 1.07, -0.53, 0.16, -0.69, -0.16. 
Standard deviation of the noise: $\sigma = 0.3$.
\item mix: length 497, changepoints at 139, 226, 243, 300, 309, 333. Values between changepoints: -0.18, 0.08, 1.07, -0.53, 0.16, -0.69, -0.16. 
Standard deviation of the noise: $\sigma = 0.3$.
\item teeth10: length 140, changepoints at 11, 21, 31, 41, 51, 61, 71, 81, 91, 101, 111, 121, 131. Values between changepoints:
 0, 1, 0, 1, 0, 1, 0, 1, 0, 1, 0, 1, 0, 1. Standard deviation of the noise: $\sigma = 0.4$.
\item stairs10: length 150, changepoints at 11, 21, 31, 41, 51, 61, 71, 81, 91, 101, 111, 121, 131, 141. Values between changepoints:
 1, 2, 3, 4, 5, 6, 7, 8, 9, 10, 11, 12, 13, 14, 15. Standard deviation of the noise: $\sigma = 0.3$.
\end{enumerate}

\end{document}